 \documentclass[longauth]{aa} 
\usepackage{graphicx}
\usepackage{txfonts}
\usepackage[]{hyperref}
\renewcommand{\arraystretch}{1.4} 
\usepackage{ulem}
\usepackage{soul}
\usepackage{xcolor}

\def\arcsec{$^{\prime\prime}$}

\def\lir{$L_{\rm IR}$\,}
\def\luv{\ifmmode L_{\rm UV} \else $L_{\rm UV}$\, \fi}

\def\Mstar{$M_{\star}$\,}

\def\micron{\ifmmode \mu\rm{m} \else $\mu$m \fi}

\def\uvbeta{$\beta$\,}

\begin{document}

   \title{The ALPINE-ALMA [CII] survey: Dust attenuation properties and obscured star formation at $z\sim4.4-5.8$}

    \titlerunning{Dust attenuation in star forming galaxies at $z\sim4.4-5.8$}

   \author{Yoshinobu Fudamoto
          \inst{1}
          \and
          P. A. Oesch\inst{1,2}
          \and
          A. Faisst\inst{3}
          \and
          M. Bethermin\inst{4}
          \and
          M. Ginolfi\inst{1}
          \and
          Y. Khusanova\inst{4}
          \and
          F. Loiacono\inst{5,6}
          \and
          O. Le F\`evre\inst{4,}\thanks{Deceased}
          \and
          P. Capak\inst{3,7,8}
          \and
          D. Schaerer\inst{1}
          \and
          J. D. Silverman\inst{9,10}
          \and
          P. Cassata\inst{11}
          \and
          L. Yan\inst{12}
          \and
          R. Amorin\inst{13,14}
          \and
           S. Bardelli\inst{6}
          \and
          M. Boquien\inst{15}
          \and
          A. Cimatti\inst{5,16}
          \and
          M. Dessauges-Zavadsky\inst{1}
          \and
          S. Fujimoto\inst{7,8}
          \and
          C. Gruppioni\inst{6}
          \and
          N. P. Hathi\inst{17}
          \and
          E. Ibar\inst{18}
          \and
          G. C. Jones\inst{19,20}
          \and
          A. M. Koekemoer\inst{17}
          \and
          G. Lagache\inst{4}
          \and
          B. C. Lemaux\inst{21}
          \and
          R. Maiolino\inst{19,20}
          \and
          D. Narayanan\inst{7,22,23}
          \and
          F. Pozzi\inst{5,6}
          \and
          D.A. Riechers\inst{24,25}
          \and
          G. Rodighiero\inst{10,26}
          \and
          M. Talia\inst{5,6}
          \and
          S. Toft\inst{7,8}
          \and
          L. Vallini\inst{27}
          \and
          D. Vergani\inst{6}
          \and
          G. Zamorani\inst{6}
          \and
           E. Zucca\inst{6}
          }

   \institute{Department of Astronomy, University of Geneva, 51 Ch. des Maillettes, 1290 Versoix, Switzerland\\
              \email{yoshinobu.fudamoto@unige.ch}
              \and
              International Associate, The Cosmic Dawn Center (DAWN)
              \and
              IPAC, California Institute of Technology,  1200 East California Boulevard, Pasadena, CA 91125, USA
              \and
              Aix Marseille Univ, CNRS, LAM, Laboratoire d'Astrophysique de Marseille, Marseille, France
              \and
              Universit{\`a} di Bologna - Dipartimento di Fisica e Astronomia, Via
                Gobetti 93/2 - I-40129, Bologna, Italy
             \and
             INAF - Osservatorio di Astrofisica e Scienza dello Spazio di
Bologna, via Gobetti 93/3, I-40129, Bologna, Italy
             \and
             The Cosmic Dawn Center (DAWN), University of Copenhagen,
Vibenshuset, Lyngbyvej 2, DK-2100 Copenhagen, Denmark
             \and
             Niels Bohr Institute, University of Copenhagen, Lyngbyvej 2, DK2100 Copenhagen, Denmark
              \and
              Kavli Institute for the Physics and Mathematics of the Universe, The University of Tokyo, Kashiwa, Japan 277-8583 (Kavli IPMU, WPI)
              \and
              Department of Astronomy, School of Science, The University of Tokyo, 7-3-1 Hongo, Bunkyo, Tokyo 113-0033, Japan
              \and
              University of Padova, Department of Physics and Astronomy  Vicolo Osservatorio 3, 35122, Padova, Italy
              \and
              The Caltech Optical Observatories, California Institute of Technology, Pasadena, CA 91125, USA
              \and
              Instituto de Investigaci{\'o}n Multidisciplinar en Ciencia y Tecnolog{\'i}a,
Universidad de La Serena, Raúl Bitr{\'a}n 1305, La Serena, Chile
              \and
              Departamento de Astronom{\'i}a, Universidad de La Serena, Av. Juan
Cisternas 1200 Norte, La Serena, Chile
             \and
             Centro de Astronom{\'i}a (CITEVA), Universidad de Antofagasta,
Avenida Angamos 601, Antofagasta, Chile
             \and
             INAF - Osservatorio Astrofisico di Arcetri, Largo E. Fermi 5, I50125, Firenze, Italy
             \and
             Space Telescope Science Institute, 3700 San Martin Drive, Baltimore, MD 21218, USA
            \and
            Instituto de F{\'i}sica y Astronom{\'i}a, Universidad de Valparaíso, Avda.
Gran Breta{\~i}a 1111, Valparaíso, Chile
            \and
            Cavendish Laboratory, University of Cambridge, 19 J. J. Thomson
Ave., Cambridge CB3 0HE, UK
            \and
            Kavli Institute for Cosmology, University of Cambridge, Madingley
Road, Cambridge CB3 0HA, UK
            \and
             Department of Physics, University of California, Davis, One Shields
Ave., Davis, CA 95616, USA
            \and
            Department of Astronomy, University of Florida, 211 Bryant Space
Sciences Center, Gainesville, FL 32611 USA
            \and
            University of Florida Informatics Institute, 432 Newell Drive, CISE Bldg E251, Gainesville, FL 32611
            \and
            Department of Astronomy, Cornell University, Space Sciences
Building, Ithaca, NY 14853, USA
            \and
            Max-Planck-Institut f{\" u}r Astronomie, K{\"o}nigstuhl 17, D-69117 Heidelberg, Germany
            \and
            INAF Osservatorio Astronomico di Padova, vicolo dellOsservatorio 5, I-35122 Padova, Italy
            \and
            Leiden Observatory, Leiden University, PO Box 9500, 2300 RA Leiden, The Netherlands
             }

   \date{Received September XX, YYYY; accepted March XX, YYYY}

 
  \abstract
    {
    We present dust attenuation properties of spectroscopically confirmed star forming galaxies on the main sequence at a  redshift of $\sim4.4-5.8$.
    Our analyses are based on the far infrared continuum observations of 118 galaxies at rest-frame $158\,\rm{\mu m}$ obtained with the Atacama Large Millimeter Array (ALMA) Large Program to INvestigate [CII] at Early times (ALPINE).
    We study the connection between the ultraviolet (UV) spectral slope ($\beta$), stellar mass ($M_{\star}$), and infrared excess (IRX$=L_{\rm{IR}}/L_{\rm{UV}}$).
    Twenty-three galaxies are individually detected in the continuum at $>3.5\,\sigma$ significance. We perform a stacking analysis using both detections and nondetections to study the average dust attenuation properties at $z\sim4.4-5.8$.
    The individual detections and stacks show that the IRX-$\beta$ relation at $z\sim5$ is consistent with a steeper dust attenuation curve than typically found at lower redshifts ($z<4$).
    The attenuation curve is similar to or even steeper than that of the extinction curve of the Small Magellanic Cloud (SMC).
    This systematic change of the IRX-$\beta$ relation as a function of redshift suggests an evolution of dust attenuation properties at $z>4$.
    Similarly, we find that our galaxies have lower IRX values, up to $1\,\rm{dex}$ on average, at a fixed mass compared to previously studied IRX-$M_{\star}$ relations at $z\lesssim4$, albeit with significant scatter. This implies a lower obscured fraction of star formation than at lower redshifts.
    Our results suggest that dust properties of UV-selected star forming galaxies at $z\gtrsim4$ are characterised by (i) a steeper attenuation curve than at $z\lesssim4$, and  (ii) a rapidly decreasing dust obscured fraction of star formation as a function of redshift. Nevertheless, even among this UV-selected sample, massive galaxies ($\log M_{\star}/M_\odot > 10$) at $z\sim5-6$ already exhibit an obscured fraction of star formation of $\sim45\%$, indicating a rapid build-up of dust during the epoch of reionization.
    }

   \keywords{Galaxies: high-redshift -- Galaxies: ISM -- ISM: dust, extinction }

   \maketitle
%

\section{Introduction}
Over the past decades, extragalactic surveys have provided large observational data of galaxies, covering wide wavelength ranges from the rest-frame ultraviolet (UV) to the rest-frame far-infrared (FIR).
These systematic, panchromatic observations have enabled the connection of the build-up of galaxies from the very early stages of their formation at high-redshift to the matured population in the local Universe.
In particular, the cosmic star formation rate density (SFRD) is found to increase at a  rapid rate until $z\sim2-3$, followed by a smooth and slow decline by an order of magnitude until the present day Universe \citep[e.g.,][]{Wilkins08,Madau2014,Bouwens2015,Oesch2018}. While great progress has been made to flesh out this overall picture, the measurement of the total SFRD at $z\gtrsim3$ is still uncertain due to uncertain dust correction factors.

The star formation rates (SFRs) of $z\gtrsim3$ galaxies are typically estimated using the UV emission from massive stars that have a short ($\sim100\,\rm{Myr}$) lifetime.
Several studies provide empirical relations to estimate the SFR directly from the UV continuum or from emission lines by atoms ionised by the UV emission \citep[e.g.,][]{Kennicutt1998,Kennicutt2012,Madau2014,Wilkins19}.
As the UV emission is highly sensitive to dust attenuation, it needs to be corrected to estimate the total intrinsic star formation activity \citep{Calzetti2000,Salim2020}.
Thus, an understanding of the dust attenuation properties as a function of redshift and other galaxy properties is one of the most important ingredients to obtain accurate SFRs, and thus to gain an accurate census of galaxy build up across cosmic history.

Absorbed UV photons heat the dust grains, which in turn re-emit the energy as thermal emission at far-infrared (FIR) wavelengths.
To correct for the absorbed UV emission, several empirical relations have been established between the dust attenuation and its FIR re-emission.
Of particular importance is the relation between the infrared excess (IRX$=L_{\rm IR}/L_{\rm UV}$) and the UV spectral slope ($\beta$: $f_{\lambda}\propto \lambda^{\beta}$). This was calibrated using local starburst galaxies \citep[e.g.,][ hereafter M99]{Meurer1999} and is routinely used in the literature to estimate the dust attenuation from high-redshift galaxies, based on the measured UV spectral colors alone.

The relation between IRX and stellar mass ($M_{\star}$) is another tool that relates the dust attenuation and stellar masses of galaxies.
The stellar mass of galaxies reflects their past star formation activity, which in turn is responsible for producing dust particles.
Therefore, the stellar mass is expected to correlate with the dust content
of the interstellar medium (ISM), as shown by observations
\citep[e.g.,][]{Heinis2014,Santiani2014,Pannella2015,Bouwens2016,Whitaker2017,Fudamoto2019,AlvarezMarquez2016,AlvarezMarquez2019}, and suggested by simulations \citep[e.g.,][]{Granzini2020}.


Both the IRX-$\beta$ and the IRX-$M_{\star}$ relations are well studied from $z\sim0$ to $z\lesssim4$, over which most authors find no significant evolution for ensemble averages \citep[e.g.,][]{Heinis2014,Pannella2015,Bouwens2016,Koprowski2018,Koprowski2020,AlvarezMarquez2016,AlvarezMarquez2019,Fudamoto2017,Fudamoto2019}.
However, at $z\gtrsim5$, these relations are still uncertain as a statistically significant sample is yet to be obtained.
While a few individual galaxies have been detected with luminous dust continuum emission even out to the highest redshifts \citep[e.g.,][]{Watson15,Laporte17,Bowler18,Hashimoto19,Tamura19}, the general trend from population averages points to rapid changes of the relation to lower IRX values at $z>5$ \citep[e.g.,][]{Capak2015,Bouwens2016,Smit18}.
In particular, using a sample of only 10 main-sequence galaxies, 
\citet{Capak2015} and \citet{Barisic2017} report a redshift evolution of the relation at $z>5$, which indicates that the IRX of $z>5$ galaxies is $\sim1\,\rm{dex}$ lower than expected from the lower-redshift IRX-$\beta$ and IRX-$M_{\star}$ relations.
If correct, it implies that the use of the ``classical'' correction would overestimate the true IR luminosity, and hence also the total SFR.
At the same time, the rapid decrease of the IR emission implies that at $z>4$ the fraction of star formation ongoing in obscured environments (i.e., the obscured fraction of star formation) becomes smaller relative to un-obscured star formation.
In previous studies, the fraction of obscured star formation as a function of stellar mass was found to be un-changed over the redshift range between $0<z<2.5$ \citep{Whitaker2017},
whereas the fainter IR emission found by \cite{Capak2015} suggests a decrease of the obscured fraction in $z>5$ star forming galaxies \citep[see also,][for theoretical predictions at $z>6$]{Wilkins2018}.
However, the existing studies at $z>4-5$ are based only on a handful ($\sim10$) of sources, and the results need to be confirmed with a larger sample size. This is the goal of the present paper.

To examine the dust attenuation properties of $z\gtrsim5$ using a large sample of normal star forming galaxies, we have studied the IRX-$\beta$, the IRX-$M_{\star}$ relation, and the obscured fraction of star forming galaxies using data from ALMA Large Program to INvestigate CII at Early Times \citep[ALPINE,][]{Lefevre2019,Bethermin2019,Faisst2019}.
ALPINE is a $\sim$70h program to observe the [CII]$158\,\rm{\mu m}$ emission lines and dust continuum of 118 normal star forming galaxies at $z\sim4.4-5.8$.
The survey provides the largest sample of normal star forming galaxies at $4<z<6$, which we here use to study the dust attenuation from a comparison
of the IR and UV emission, and to examine how it relates with observable and derived galaxy properties.

This paper is organised as follows: in \S2 we describe our sample and observations, and in \S3 we presents the methods of our analyses.  \S4 shows the results on the IRX-$\beta$/$M_{\star}$ relations and the obscured fraction of star formation obtained from our sample. Finally, we conclude our study in \S5.
Throughout this paper, we assume a cosmology with $(\Omega_m,\Omega_{\lambda},h)=(0.3,0.7,0.7)$, and the Chabrier \citep{Chabrier2003} initial mass function (IMF) where applicable.

\section{Observations}
In this section, we briefly discuss the ALPINE galaxy sample and observations that are relevant to this study. We refer to \citet{Lefevre2019}, \citet{Bethermin2019}, and \citet{Faisst2019} for a complete description of the survey objectives, the ALMA data processing, and the multiwavelength ancillary observations, respectively.

\subsection{Sample and ancillary data}


ALPINE observed 118 star forming galaxies at  $z\sim 4.4-5.9$ \citep[see][]{Lefevre2019}.
The targets have secure spectroscopic redshifts from two observation campaigns in the COSMOS (105) and the GOODS-South (13) fields using rest-frame UV emission and/or absorption lines \citep[][]{Lefevre2015,Haisinger2018}.

Using the vast amount of ancillary data available for these fields, 
we performed SED fitting to estimate SFRs, stellar masses, UV luminosities, UV spectral slopes, and other quantities, as described in detail by \citet{Faisst2019}.
We adopt these SED-based quantities from \citet{Faisst2019} for this paper.

Two galaxies in the sample are confirmed AGN based on an X-ray detection \citep[DEIMOS\_COSMOS\_845652;][]{Faisst2019} and from deep optical spectroscopy \citep[CANDELS\_GOODS\_14;][]{Grazian2020}.
As AGNs could outshine the rest-UV emission of stars, we removed these two sources from further analysis. 
Additionally, a stack of all the Chandra X-ray images of the remaining galaxies showed no detection, confirming that our sample is not significantly AGN-contaminated, on average. 
While heavily dust-obscured AGNs are not excluded by these rest-UV and X-ray criteria above, they should have little impact on the FIR emission probed by our ALMA data which is dominated by relatively cold dust.

\begin{figure}[tb]
    \centering
    \includegraphics[width=\columnwidth]{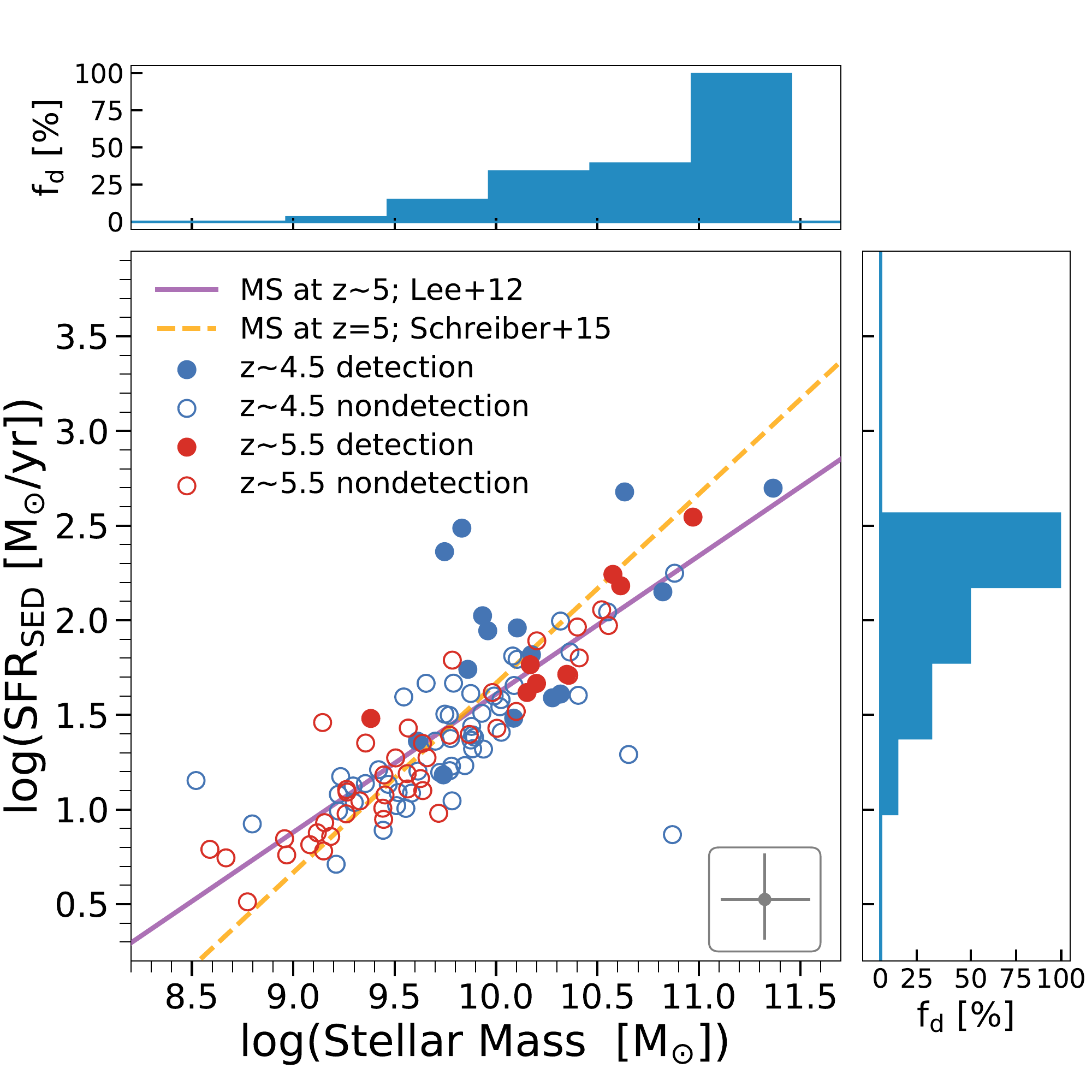}
    \caption{
    Stellar mass and SFR diagram of our galaxies estimated using the SED fitting code LePhare \citep{Ilbert2006,Arnouts1999}.
    The top and the right panels show detection rates ($f_{\rm d}$) of continuum emission as functions of SFR and stellar mass.
     Filled and open points in the middle panel represent the continuum detected ($>3.5\,\sigma$) and the non-detected galaxies (blue: $z\sim4.5$, red: $z\sim5.5$ galaxies), respectively.
    Solid and dashed lines show two different estimates of the $z\sim5$ main-sequence of star forming galaxies \citep{Lee2012,Schreiber2015}.
    Typical uncertainty of SFR and stellar mass are shown in the bottom right inset.
    The continuum detected galaxies mostly show stellar mass above $\sim10^{10}\,\rm{M_{\odot}}$, and SFR above $\sim30\,\rm{M_{\odot}/yr}$.
    }
    \label{fig:MS_detection}
\end{figure}

By selection, the galaxies from the ALPINE survey have SFRs and stellar masses, which are consistent with the main-sequence of star formation at $z\sim5$ \citep[see Fig. \ref{fig:MS_detection};][]{Steinhardt2014,Schreiber2015,Faisst2019}.
The ranges of stellar mass and total SFR spanned by our targets are $\rm{log\,(M_{\star}/M_{\odot}) \sim 8.5 - 11.5}$ and $\rm{log\,(SFR/M_{\odot}\,yr^{-1}}) \sim 0.5 - 2.5$, respectively.

As part of the SED fitting procedure in \citet{Faisst2019}, we calculated monochromatic UV luminosities, $L_{\rm UV}$, without dust attenuation corrections at rest frame $1600$\AA, and fitted a power-law function ($f_{\lambda}\propto\lambda^{\beta}$) to measure the UV continuum slopes $\beta$ using the wavelength range 1300\AA\,to 2300\AA, consistent with previous studies \citep[e.g.,][]{Bouwens2016,Mclure2018,Fudamoto2019}.
Uncertainties were estimated from Monte Carlo simulations, by perturbing the fluxes of each filter assuming a Gaussian error distribution in the photometric measurement.
We used median values as our best fits, and 16th and 84th percentiles as lower and upper uncertainties of the \luv and $\beta$ measurements.

Based on the \luv measurements, we calculated UV-based SFRs without dust attenuation corrections by employing the equation from \citet{Madau2014} converted to a Chabrier IMF as follows
\begin{equation}
    \label{eqn:sfr_uv}
	\rm{SFR_{UV}\,(M_{\star}\,\rm{yr^{-1}})} = 0.76\times10^{-28}\,L_{\rm{UV}}\,(\rm{erg\,s^{-1}\,Hz^{-1}}).
\end{equation}
The above conversion provides consistent SFRs with other studies \citep[e.g.,][]{Kennicutt1998}, within our \luv\,measurement errors.

\subsection{ALMA observations and data reduction}
\label{sec:observations}
The details of the ALMA observations and the data reduction are presented in \citet{Bethermin2019}.
In short,
ALMA observed our sample between 07 May, 2018 (Cycle 5) and 10 January, 2019 (Cycle 6) using  antenna configurations C43-1 and C43-2.
The lower side bands were used for both continuum measurements and  the [CII] $158\,\rm{\mu m}$ emission line expected from the redshifts, and the upper side bands covered the dust continuum only.
The integration times ranged from 15 to 45 minutes, with an average of 22 minutes.


After a basic calibration using the pipeline of the Common Astronomy Software Applications package \citep[CASA;][]{Mcmullin2007}, and additional flagging for bad antennae, 
continuum maps were produced using the line free channels.
In particular, we excluded channels within the $\pm3\,\sigma$ width of the detected [CII] emission lines (i.e., excluding above $\sim1\%$ of the maximum amplitude assuming a Gaussian profile).
When the [CII] emissions have complex morphology \citep[e.g., mergers or potential outflow;][]{Jones2019,Ginolfi2019}, the $\pm3\,\sigma$ widths may not exclude all components of the [CII] lines.
In these cases, we further extended the [CII] line masks by $\sim0.1-0.2\,\rm{GHz}$ to prevent the [CII] line from contaminating our continuum maps.
Using these [CII] masks, we imaged continuum maps using the CASA task {\fontfamily{cmtt}\selectfont{TCLEAN}} with the natural weighting scheme to maximise sensitivity.

The resulting median (minimum - maximum) point-source sensitivities and  resolutions are $41\, \rm{\mu Jy/beam}$ ($16.8-72.1\, \rm{\mu Jy/beam}$) and 1.1\arcsec ($0.9-1.6^{\prime\prime}$), respectively.

\subsection{FIR continuum measurements and detection fractions}
\label{sec:photometry}

As described in \citet{Bethermin2019},
the detection threshold used for the FIR continuum is $3.5\,\sigma$ within 2\arcsec\,diameters of the UV counterpart position.
Signal-to-noise ratios were determined using peak pixels and background RMS.
With this conservative approach, the fidelity of our detections is $>95\%$ \citep{Bethermin2019}. 
In total, 23 of our galaxies have continuum detections at this level.
We measured flux densities using 2D Gaussian fitting with our customised routine.
Flux measurement uncertainties were estimated using a method provided by \citet{Condon1997}, which properly accounts for correlated noise as present in interferometric data.

For nondetections, we used upper limits for our analyses.
We determined $3\,\sigma$ upper limits by searching the maximum flux within 2\arcsec\ diameter of the UV counterpart position, and by adding three times the background RMS to that maximum value.
These upper limits are more conservative than in most previous works.
In particular, these values account for potentially weak continuum signals that are just below the detection threshold \citep[see][for a discussion]{Bethermin2019}.

Figure \ref{fig:MS_detection} summarizes the continuum detections as a function of stellar mass and SFR. Clearly, the detection fractions are strongly mass and SFR dependent.
Most of the FIR detected galaxies are massive ($M_{\star} \gtrsim 10^{10}\,\rm{M_{\odot}}$),
and highly star forming ($\rm{SFR_{SED} \gtrsim 30\,\rm{M_{\odot}/yr}}$) galaxies.
Nevertheless, some galaxies at the massive and the most star forming end did not show significant continuum detections, even though the sensitivity does not largely change.
This suggests that the $\rm{SFR_{SED}}$ and/or stellar mass alone are not ideal indicators of \lir.

\section{Analysis}
\label{sec:analysis}


\subsection{\lir\,estimation from ALMA observations}

\label{sec:lirest}
The total infrared luminosities \lir\,over the wavelength  range of $\lambda_{\rm{rest}}=8-1000\,\rm{\mu m}$ were estimated using the conversion factor from $158\,\rm{\mu m}$ to \lir presented in \citet{Bethermin2019}. In particular, \citet{Bethermin2019} constructed a mean stacked continuum FIR SED for ALPINE galaxy analogs in terms of redshift, stellar mass and SFR, using the full multiwavelength dataset available in the COSMOS field \citep{Davidzon2017}. This includes deep {\it Herschel} \citep{Lutz2011,Oliver2012}, AzTEC/ASTE \citep{Aretxaga2011}, and SCUBA2 data \citep{Casey2013}. The mean stacked SED was then fit with a FIR template using the \citet{Bethermin2017} model, which uses an updated version of \citet{Magdis2012} templates based on the evolution of the average dust SEDs.
The template has a UV interstellar radiation field $\langle U \rangle=50$, as parameterised in \citet{Bethermin2015}.
In particular, the template matches a modified black body with a relatively high peak dust temperature ($T_{\rm{d}}\sim41\,\rm{K}$) with an additional mid-infrared component that reproduces the {\it Herschel} fluxes.  


This relatively high dust temperature is consistent with the extrapolation of lower redshift trends (e.g., $T_{\rm{d}}>40\,\rm{K}$ for $z\gtrsim3$), as reported in several studies  \citep{Bethermin2015,AlvarezMarquez2016,Ferrara2017,Faisst2017,Schreiber2018,Fudamoto2019}.
We note, however, that we do not have constraints on the FIR SEDs of individual galaxies. If extreme variations of dust temperatures are present within our sample, this will thus introduce a scatter on the derived ALMA continuum to \lir\,conversion factor.
Since we are interested in the population average \lir\ values, however, we can ignore this scatter and simply adopt the best-fit template to the mean stacked FIR SED as derived in \citet{Bethermin2019}.


To derive the \lir\ values in practice, our template was normalised to the continuum fluxes or $3\,\sigma$ upper limits measured at rest-frame $158\,\rm{\mu m}$, and then integrated over the wavelength range $8-1000\,{\rm \mu m}$.
In this way, the measured monochromatic luminosity at $\lambda_{\rm rest}=158\,\rm{\mu m}$ was converted to \lir\,by $\nu_{158\,\rm{\mu m}}\,L_{\nu_{158\,\rm{\mu m}}}/L_{\rm{IR}}=0.13$ \citep{Bethermin2019}. 
The uncertainties on \lir\,were derived directly from the flux measurement uncertainties.
Based on this \lir\,and the attenuation uncorrected \luv\,derived from the optical photometry, we computed the infrared excess as IRX$=L_{\rm IR}/L_{\rm UV}$, with the proper uncertainties.

Finally, we converted the \lir\,to obscured star formation rates by employing the equation from \citet{Madau2014} converted to the Chabrier IMF as follows
\begin{equation}
\label{eqn:sfr_ir}
\rm{SFR_{IR}\,(M_{\star}\,\rm{yr^{-1}})} = 2.64\times10^{-44}\,L_{\rm{IR}}\,(\rm{erg\,s^{-1}}),
\end{equation}
which is consistent with the values derived in \citet{Kennicutt1998}.

\subsection{Stacking analysis}
\label{sec:stacking}
To obtain average properties of our galaxies, we performed a stacking analysis of all ALMA continuum images, including both individual detections and nondetections.
We used image based stacking as described in Khusanova (in preparation), which we briefly summarise here.

We performed stacking using the $\lambda_{\rm{rest}}=158\,\rm{\mu m}$ continuum images centered on the UV counterpart positions.
To create stacked images free from projected bright sources, we masked all serendipitous sources detected above $5\,\sigma$ significance based on the serendipitous detection catalog from \citet{Bethermin2019}.
We used weighted median stacking. That is, after aligning all images to the UV-based phase center, the stacked images were  constructed by comparing the distributions of intensities for each pixel and taking a weighted median as follows:

\begin{align}
\label{eqn:stack}
\begin{split}
 f_{\nu_{\rm 158\mu m}}^{\rm Median} &= {\rm Median}(f_{\nu_{\rm158\mu m}\,,i}\,w_i) ,
\\
 w_i &= \frac{L_{\rm UV}^{\rm Median}}{L_{{\rm UV}, i}} ,
\end{split}
\end{align}
where $f_{\nu_{\rm 158\mu m}}^{\rm Median}$ is the weighted median flux, $f_{\nu_{\rm158\mu m}\,,i}$ is the rest-frame $158\,\rm{\mu m}$ continuum flux of the i-th galaxy image, $L_{\rm UV}^{\rm Median}$ is the median UV luminosity of all galaxies in the stacking bin, and $L_{\rm UV, i}$ is the UV luminosity of the i-th galaxy. The inverse \luv weighting scheme was chosen to provide us with an accurate measure of the IRX (\lir/\luv). For testing purposes, we also computed unweighted median stacks, which did not differ significantly from our weighted medians. 

If the UV and IR components within a given galaxy are significantly offset spatially, it could be possible that we lose part of the IR fluxes in our stacking procedure, as we use the UV-based phase center as the stacking reference. To test this, we performed stacks using the individual FIR continuum detections, and found that both the UV-centered stacks and the FIR-centered stacks resulted in identical stacked fluxes. This is consistent with the finding in \citet{Fujimoto2020}, who concluded that the UV and FIR continuum positions of our sample are not significantly offset within the relatively large beam size ($\sim1^{\prime\prime}$) of our observations.


Flux densities were measured from the stacks using aperture photometry ($r=1.5^{\prime\prime}$), and the measurement uncertainties were estimated by boot strapping.
As for our individual continuum detections (cf.\ Sect.\ \ref{sec:photometry}), our detection thresholds for stacks are $3.5\,\sigma$ using the peak pixel flux density, where $\sigma$ is the pixel background RMS.
When no continuum is detected in the stacks at this level, we determined conservative $3\,\sigma$ upper limits by searching the maximum pixel value within $2^{\prime\prime}$ of the expected position, and by adding three times the background RMS to the local maximum, as done for the individual sources in the ALPINE catalog \citep{Bethermin2019}.

\begin{figure}[tb]
    \centering
    \includegraphics[width=3.3in]{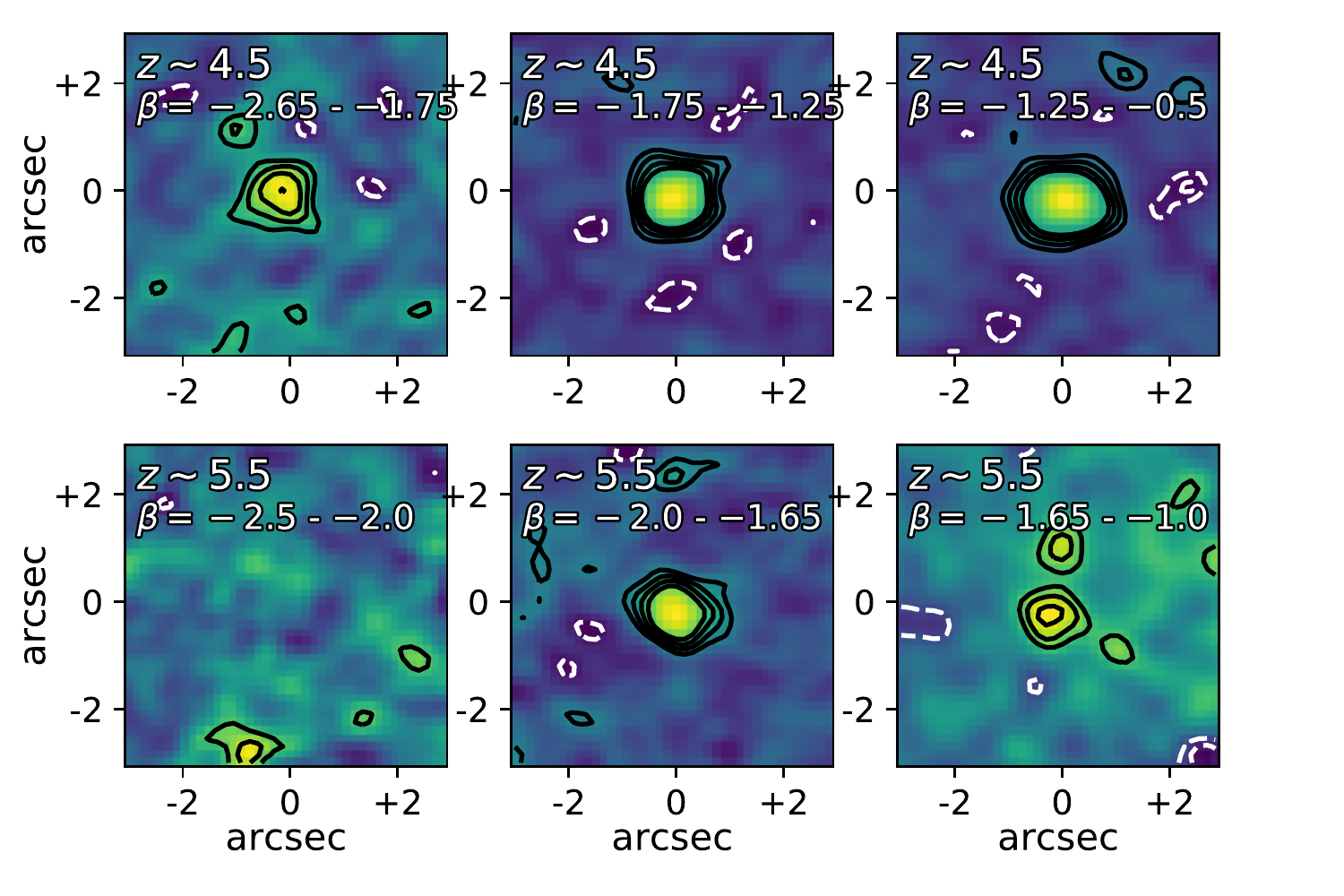}
    \caption{
    6\arcsec$\times$6\arcsec\,cutouts of $\beta$ binned stacks of ALMA continuum images used to derive the stacked infrared luminosities.
    The upper and lower panels show stacks of galaxies at $z\sim4.5$ and $z\sim5.5$, respectively.
    Black solid contours show $2,3,4,5\,\sigma$ and white dashed contours show $-3,-2\,\sigma$, if present.
    The ranges of $\beta$ used in each stacks are shown in each cutouts.
    While in the $z\sim4.5$ bins all stacks have clear detections at $>3.5\,\sigma$, 
    in the $z\sim5.5$ bins the stack in the bluest bin (lower left panel) remains undetected and we only report a conservative $3\sigma$ upper limit (see text for details).
    }
    \label{fig:betastack}
\end{figure}

\begin{figure}[tb]
    \centering
    \includegraphics[width=0.76\columnwidth]{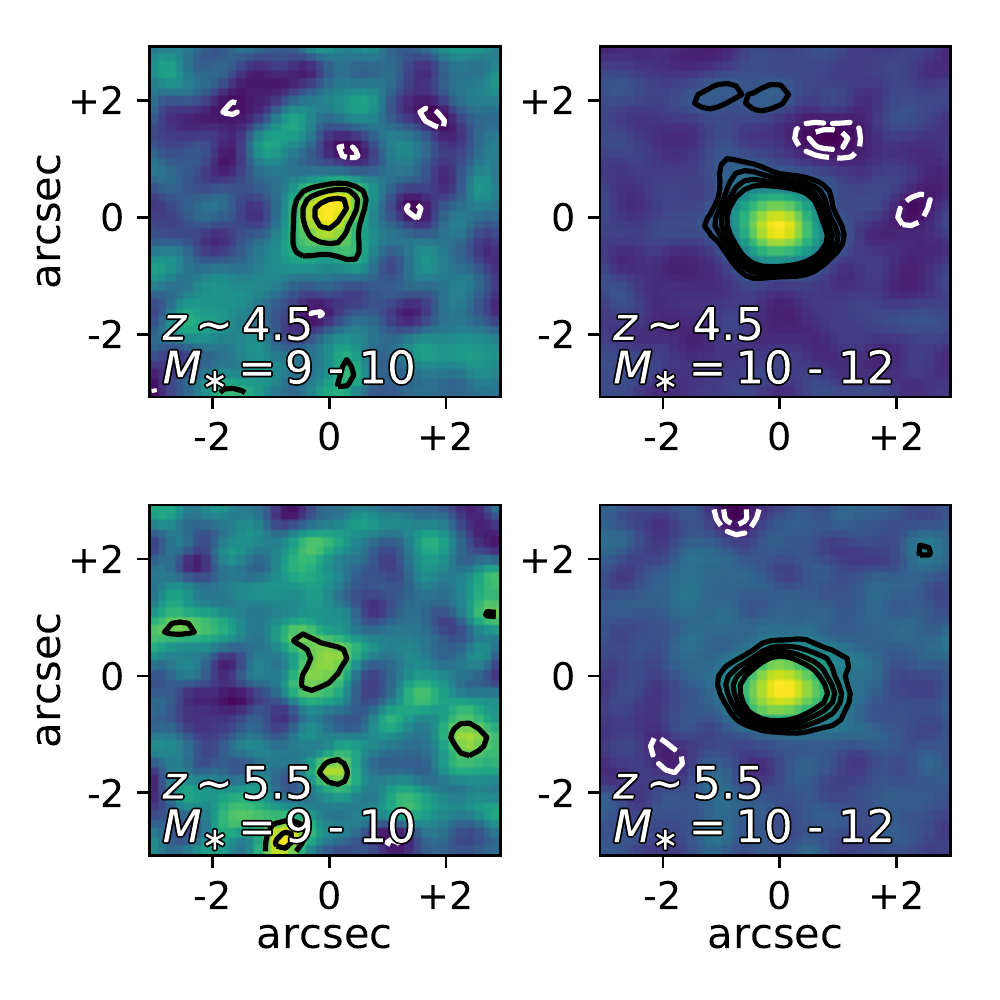}
    \caption{
    Same as Figure \ref{fig:betastack}, but for ${M_{\star}}$ binned stacks.
    The ranges of ${M_{\star}}$ used in each stacks are shown in each cutouts (in log solar masses).
    While we obtained strong detections from most of the stacked images,
    in the lower mass bin of $z\sim5.5$ galaxies, only a $\sim2\,\sigma$ peak is present at the source position. This is lower than our detection threshold for individual images, therefore we provide a conservative $3\,\sigma$ upper limit for this bin (see text).
    }
    \label{fig:massstack}
\end{figure}

We performed stacking in two different redshift bins of $z<5$ and $z>5$.
In each redshift bin, we further split the sample in bins based both on UV slope and on stellar mass.
For the $\beta$, $L_{\rm UV}$, and $M_{\star}$ values of each bin, we used the median values of the appropriate sample.
The results of our stacking analysis are summarised in  Table \ref{tab:stack}, while the stacked images are shown in Figures \ref{fig:betastack} and \ref{fig:massstack}.

For the IRX-$\beta$ analysis
the bins were chosen to split the sample roughly equally in the two different redshift bins. Specifically, at $z<5$, we used $\beta=[-2.65,-1.75]$, $[-1.75,-1.25]$, $[-1.25,-0.5]$, while for $z>5$ galaxies we used $\beta=[-2.65,-2.0]$, $[-2.0,-1.7]$, $[-1.7,-1.0]$ (Fig. \ref{fig:betastack}).

From the stacks at $4<z<5$, we detected significant continuum emission from all $\beta$ bins, while from the galaxies at $5<z<6$ we detected all but the bluest bin (i.e., $\beta=-2.65$ - $-2.0$).
We note that the stellar mass range used in this $\beta$ binning is comparable with the mass range used in several previous studies, such that our results can be directly compared \citep{AlvarezMarquez2019,Fudamoto2017,Fudamoto2019}.

Stacks based on $M_{\star}$ bins were used for the IRX-$M_{\star}$ analysis binned by $\rm{log}(M_{\star}/\rm{M_{\odot})}=[9,10]$ and $\rm{log}(M_{\star}/\rm{M_{\odot})}=[10,12]$ for both $z<5$ and $z>5$ galaxies.  In total, 111 galaxies are used, while a few very low mass galaxies (7) are not included in these stacks.

From the  stacks at $4<z<5$, we detected $158\,\rm{\mu m}$ continuum from all $M_{\star}$ bins. In the range $5<z<6$, the stack in the high mass bin has a strong detection, while the lower mass bin only shows a tentative signal at $\sim3\,\sigma$, which we report as a nondetection and provide our conservative $3\,\sigma$ upper limit (Fig. \ref{fig:massstack} and Table \ref{tab:stack}).

\begin{table}
	\begin{center}
	\caption{Results of the Stacking Analysis }
	\label{tab:stack}
	\begin{tabular}{cccccc}
		\hline\hline
		redshift$^{\rm{a}}$&\uvbeta$^{\rm{a}}$ & \# of sources & log \Mstar$^{\rm{a}}$&log \lir& logIRX\\
		&&&[$M_{\sun}$]&[$L_{\sun}$]&\\
		\hline
		\multicolumn{5}{c}{$\beta$ Stacks}\\[2pt]
		4.53&-2.00&34&9.71&$10.91^{+0.16}_{-0.20}$&$-0.09^{+0.16}_{-0.22}$\\
		4.53&-1.47&24&9.91&$11.01^{+0.18}_{-0.24}$&$0.20^{+0.17}_{-0.29}$\\
		4.52&-0.86&8&10.37&$11.67^{+0.09}_{-0.15}$&$0.69^{+0.11}_{-0.12}$\\[5pt]
		5.68&-2.28&22&9.23&$<$10.77&$<$-0.03\\
		5.57&-1.84&14&9.98&$10.89^{+0.16}_{-0.13}$&$-0.15^{+0.16}_{-0.13}$\\
		5.51&-1.43&11&10.35&$11.15^{+0.11}_{-0.25}$&$0.12^{+0.17}_{-0.38}$\\
		\hline
		\multicolumn{5}{c}{$M_{\star}$ Stacks}\\[2pt]
		4.53&-1.71&35&9.74&$10.74^{+0.21}_{-0.41}$&$-0.09^{+0.23}_{-0.44}$\\
		4.54&-1.36&28&10.24&$11.36^{+0.11}_{-0.11}$&$0.37^{+0.10}_{-0.14}$\\[5pt]
		5.66&-2.08&32&9.61&$<$10.85&$<$0.04\\
		5.54&-1.51&14&10.46&$11.18^{+0.23}_{-0.20}$&$0.05^{+0.19}_{-0.21}$\\
		\hline
	\end{tabular}
	\end{center}
	$\rm^{a}$ Median values in each bin.
	\vspace{4pt}
	\end{table}

\section{Results}
\label{sec:results}

\begin{figure*}
    \centering
    \includegraphics[width=7.3in]{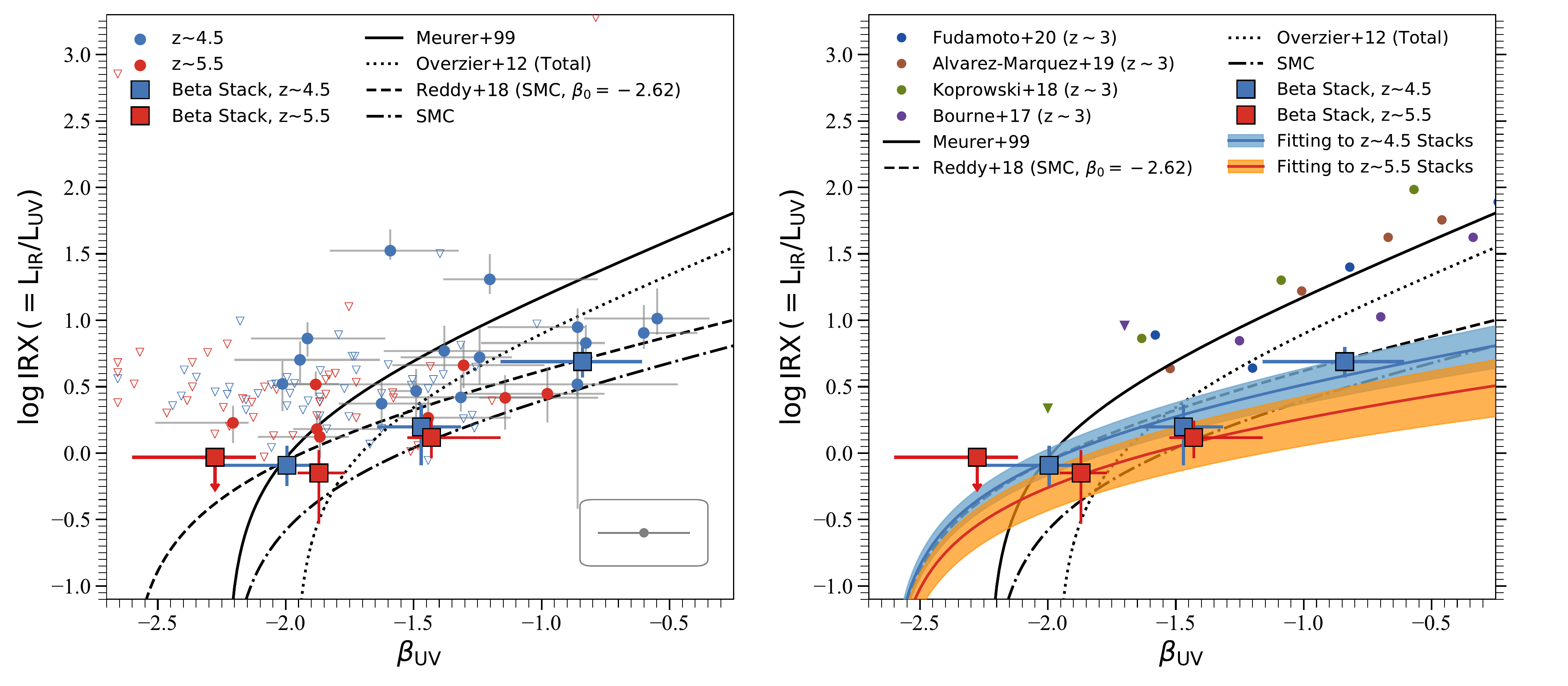}
    \caption{
    {\bf Left}: IRX-\uvbeta\,diagram of ALPINE galaxies with previously determined relations at two different redshifts.
    Blue and red points show individual FIR continuum detections at $4<z<5$ and at $5<z<6$, respectively.
    Open downward triangles show $3\,\sigma$ upper limits of individual IR nondetections.
    Stacks are shown as blue (at $z\sim4.5$) and as red (at $z\sim5.5$) rectangles.
    The nondetection of stack is indicated by a downward arrow.
    In the bottom right of the left panel, a horizontal bar shows the median uncertainty of individual \uvbeta\,measurements ($\Delta\beta=\pm0.18$). 
    Solid and dotted lines show the IRX-\uvbeta relation of local starbursts from \citet{Meurer1999} and that assuming SMC dust attenuation \citep[e.g.,][]{Prevot1984}, respectively.
    We also plot the updated local relation from \citet{Overzier2011} (dotted line).
    Assuming bluer intrinsic $\beta$, \citet{Reddy2018} proposed an IRX-$\beta$ relation at $z\sim2$ which has a SMC dust extinction curve and bluer $\beta$ (dot-dashed line).
    {\bf Right}: Fitting results to the stacks at each redshift assuming an intrinsic $\beta_0$ of -2.62 are shown with red and blue lines with $1\sigma$ uncertainty (orange and blue bands). Small points and downward  triangles show representative stacking detections and upper limits of $z\sim3$ galaxies, respectively \citep{Bourne2017,Koprowski2018,AlvarezMarquez2019,Fudamoto2019}.
    Both at $z\sim4.5$ and $z\sim5.5$, our stacks show lower IRX than $z\sim3$ results at fixed $\beta$. The IRX-$\beta$ relation from \citet{Meurer1999} is not consistent with our results. Within uncertainties, at $z\sim4.5$ and at $z\sim5.5$, our sample prefers the IRX-$\beta$ relation from \citet{Reddy2018}. Our stacks and fitting results indicate that galaxies follow an IRX-$\beta$ relation most similar to an SMC type of attenuation.
    }
    \label{fig:IRX_ALPINE}
\end{figure*}

\subsection{IRX-$\beta$ relation}
\label{sec:IRX-beta}

We study the IRX-$\beta$ relations by separating our sample in two different redshift intervals, at $4<z<5$ and $5<z<6$. We then compare our results with existing studies at $z<4$ to investigate the evolution of dust attenuation properties over a wide redshift range between $z\sim0$ and $z\sim6$. 

In order to connect the IRX-$\beta$ diagram to the attenuation properties for an ensemble of galaxies, one has to make an assumption about their intrinsic, dust-free UV continuum slope $\beta_0$, which depends on stellar population properties (e.g., metallicity, age). Once this is fixed, the slope of the attenuation curve directly determines the position of galaxies in the IRX-$\beta$ space \citep[see e.g.,][]{Salim2020}.
In the following, we parameterize the slope of the attenuation curve as the change in FUV attenuation for a given change in UV continuum slope, $d\,A_{\rm FUV}/d\beta$, following previous analyses \citep[e.g.,][]{Meurer1999, Bouwens2016}.
The IRX-$\beta$ relation can then be written as
\begin{equation}
    \label{eqn:irxb}
	\rm{IRX} =  1.75 \times \left( 10^{0.4\frac{d\,A_{\rm FUV}}{d\beta}(\beta - \beta_0)} -1 \right),
\end{equation}
where the prefactor 1.75 comes from the bolometric correction of the UV luminosity (${BC_{\rm UV}}$). Several studies derived $BC_{\rm UV}\sim1.7$ \citep[e.g.,][]{Meurer1999, Bouwens2016, Koprowski2018}, and here we used $BC_{\rm UV}=1.75$ to make our assumption on the IRX-$\beta$ relation consistent with the recent ALMA based study \citep{Bouwens2016}. 
We refer to ``Meurer-like'' and ``SMC-like'' attenuation based on the UV reddening slopes of $d\,A_{\rm FUV}/d\beta = 1.99$ \citep{Meurer1999} and $d\,A_{\rm FUV}/d\beta = 1.1$, respectively \citep[see also][]{Reddy2018}. Note that we are thus implicitly treating the extinction curve as measured toward stars in the SMC as an attenuation curve.

\subsubsection{The observed IRX-$\beta$ relation}

The measured IRX-$\beta$ diagram of our $z\sim4-6$ galaxy sample is shown in Figure \ref{fig:IRX_ALPINE}. 
From the overall distribution of all the individual detections, $3\sigma$ upper limits, and stacks, we find that our galaxies generally do not follow the Meurer-like IRX-$\beta$ relations from local starbursts, or from $z\sim3$ massive ($M_{\star}\gtrsim10^{10}\,\rm{M_{\odot}}$) star forming galaxies \citep[ e.g.,][]{Bouwens2019,AlvarezMarquez2019,Fudamoto2017,Fudamoto2019}.
The local IRX-$\beta$ relation updated with photometry using larger apertures \citep[i.e.,][]{Overzier2011,Takeuchi2012} are more consistent with our individual detections. However, these IRX-$\beta$ relations require a relatively red or large $\beta_0$ \citep[e.g., $\beta_0=-1.96$,][]{Overzier2011}, which is not consistent with many of our $\beta$ measurements including nondetections and the theoretically predicted evolution of $\beta_0$ of metal poor high-redshift galaxies \citep[e.g.,][]{Wilkins2011,Alavi2014,Reddy2018}.

The lower IRX values of our sample suggest an evolution of the average dust attenuation properties at $z>3$, becoming more similar to an SMC-like dust attenuation. However, we note the very large dispersion present in IRX values at fixed $\beta$ for individual detections, which reaches more than 1 dex relative to the stacked values in some cases. Such scatter to higher IRX has been seen before, and can be explained with geometric effects  \citep[e.g.,][]{Howell2010,Casey2014,Popping2017,Narayanan2018,Faisst2019}. Despite this scatter within the population, it is clear that the average IRX values at fixed $\beta$, as measured through our stacks, are evolving compared to lower redshift derivations. 


In detail, at $z\sim4.5$, more than half of the galaxies which are individually detected (10 out of 15) are located below the M99 relation. 
Furthermore, all the stacks with $\beta>-1.75$ lie $\gtrsim0.5\,{\rm dex}$ below the M99 relation. Based on these results, we conclude that the  IRX-$\beta$ relation of our sample at $z\sim4.5$ does not follow a Meurer-like IRX-$\beta$ relation, but requires a steeper attenuation curve.

At $z\sim5.5$, we detected only a small fraction of galaxies in the dust continuum (8 out of 51), but the majority of these individually detected sources lie somewhat below the M99 relation.
Although the stack of the redder galaxies ($\beta=[-1.75,-1.0]$) is perfectly consistent with an SMC-like IRX-$\beta$ relation, the stack of bluer galaxies ($\beta=[-2.0,-1.75]$) lies slightly above the SMC-like relation, while the bluest bin only resulted in an upper limit on the IRX. Nevertheless, we conclude that an SMC-like IRX-$\beta$ relation is more consistent with our $z\sim5.5$ galaxy sample than the Meurer-like relation, as indicated also by other studies \citep{Capak2015,Barisic2017}. We quantify this in detail in the next section.

%

\subsubsection{Fitting the IRX-$\beta$ relation}
Given the stacked IRX values as a function of $\beta$ derived above, we can quantify the slope of the attenuation curve of our high-redshift galaxies by fitting the IRX-$\beta$ relation. As shown in Equation \ref{eqn:irxb}, the IRX-$\beta$ relation depends both on the shape of the dust curve, through $d\,A_{\rm FUV}/d\beta$, and on the intrinsic UV continuum slope $\beta_0$. While, the classical value $\beta_0=-2.23$ derived in M99 for local galaxies is heavily used in the literature also for higher-redshift sources, several authors have pointed out that the physical conditions in early galaxies are likely very different, which affect the intrinsic slope, including younger stellar populations, lower metallicities, or potential changes to the initial mass function \citep[e.g.,][]{Wilkins2011,Alavi2014,Castellano2014}. In particular, \citet{Reddy2018} derived a more appropriate value of $\beta_0=-2.62$ for early galaxies, based on the binary population and spectral synthesis model \citep[BPASS;][]{Eldridge2012,Stanway2016} with a stellar metallicity of $Z=0.14\,Z_{\odot}$. 

Indeed, this bluer $\beta_0$ is clearly more consistent with the $\beta$ distribution of our sample, which includes several galaxies that are significantly bluer than the nominal dust-free value of M99. In the following, we thus use $\beta_0=-2.62$ as our baseline value for our fits of the IRX-$\beta$ relation, but we  discuss how our results change by using the classical value of $\beta_0=-2.23$ derived in M99.


Using the IRX-$\beta$ relation (Equation \ref{eqn:irxb}) with fixed $\beta_0=-2.62$, we fit the stacked IRX values as a function of $\beta$, which results in $d\,A_{\rm FUV}/d\beta = 0.71\pm0.15$ at $z\sim4.5$ and $d\,A_{\rm FUV}/d\beta = 0.48\pm0.13$ at $z\sim5.5$ (see Figure \ref{fig:IRX_ALPINE} and Table \ref{table:IRXb}). This is significantly steeper than a Meurer-like attenuation of $d\,A_{\rm FUV}/d\beta = 1.99$, meaning that less UV attenuation is required to result in a given reddening of the UV slope.  
Therefore,
in both redshift bins, our data reject a Meurer-like attenuation curve at $>5\,\sigma$. The best-fit values even lie below the SMC extinction curve of $d\,A_{\rm FUV}/d\beta = 1.1$, as can be appreciated from the comparison of our best-fit curves in Fig. \ref{fig:IRX_ALPINE} with the line derived in \citet{Reddy2018} using the same $\beta_0$.

Our current data are unfortunately not good enough to constrain $\beta_0$ of our sample, as such constraints require extremely deep IR observations of blue galaxies with accurate $\beta$ measurements.
However, the exact $\beta_0$ value does not affect our conclusion of the SMC-like dust properties at $z>4$. In particular, even if we change $\beta_0$ to the classical value from M99 of $\beta_0=-2.23$, we still derive a best-fit $d\,A_{\rm FUV}/d\beta = 1.01\pm0.20$ for $4<z<5$ galaxies and $d\,A_{\rm FUV}/d\beta = 0.74\pm0.24$ values for $5<z<6$ galaxies (see Table \ref{table:IRXb}). These values also exclude a Meurer-like attenuation at $\sim4\,\sigma$ and at $\sim5\,\sigma$, respectively. Thus, our observations indicate that the attenuation curve of $z>4$ UV-selected main-sequence galaxies are SMC-like, irrespective of the intrinsic UV slopes.
In the appendix, we summarise the predicted
IR luminosities based on our best fit IRX-$\beta$ relations in Table \ref{table:A_LIR}, together with the measured $L_{\rm{IR}}$ for detections.


\begin{table}
\caption{Fitting Results of the IRX-$\beta$ Relation}
\label{table:IRXb}
\centering
\begin{tabular}{c  c  c}
\hline\hline                 
Redshift Bin  & $d\,A_{\rm FUV}/d\beta$ &  $d\,A_{\rm FUV}/d\beta$\\ 
  & assuming $\beta_0=-2.62$ &  assuming $\beta_0=-2.23$ \\ 
\hline

 4.5 &  $0.71\pm0.15$ & $1.01\pm0.20$ \\

5.5 &   $0.48\pm0.13$ & $0.74\pm0.24$\\
\hline
\end{tabular}
\end{table}

As shown in Figure \ref{fig:IRX_ALPINE}, previous studies found that the IRX-$\beta$ relations of UV selected star forming galaxies at $z\sim3-4$ are consistent with the M99 relation using stacks of {\it Herschel} images \citep{Heinis2014,AlvarezMarquez2016,Koprowski2018,AlvarezMarquez2019}.
Also from individual detections and stacking analyses of ALMA observations, studies showed that massive UV selected, star forming galaxies at $z\sim3-4$ are consistent with the M99 relation, while lower mass galaxies show a relation close to the SMC-like IRX-$\beta$ \citep{Bouwens2016,Fudamoto2019}.
These previous $z\sim3-4$ observations are based on UV selected star forming galaxies (such as Lyman Break galaxies) similar to our analysis here, and they included both detections and nondetections in their stacks. Hence, they should be directly comparable to our work. Nevertheless, 
at the higher redshift probed by our sample here, our results indicate that SMC type relations are more applicable even for massive galaxies.
Even within our sample itself, we find tentative evidence for a redshift evolution of the attenuation properties, given the different values derived for $d\,A_{\rm FUV}/d\beta$ at the $2\sigma$ level between $z\sim4.5$ and $z\sim5.5$.
This is consistent with the previous analysis at $z\sim5.5$ \citep{Capak2015,Barisic2017}.
Overall, these results indicate an evolution of the average dust properties (such as grain size distribution and/or composition) of UV-selected main-sequence galaxies at $z\gtrsim3$.



\subsubsection{Systematic trends with morpho-kinematics}
In addition to the overall relation, we investigated if there is any correlations between a galaxy's location in the IRX-$\beta$ diagram and its morphology and/or kinematic conditions. Based on the detected [CII] emission lines kinematics and morphology, our sample was classified as rotators (9 galaxies), mergers (31 galaxies), extended dispersion dominated (15 galaxies), and compact dispersion dominated (8 galaxies) galaxies \citep{Lefevre2019}.
We did not find any clear trends that systematically correlate with the locations of galaxies in the IRX-$\beta$ diagram. Nevertheless, we noted that  none of the compact, dispersion dominated galaxies ($\lesssim10\,\%$ of our whole sample) are individually detected in the continuum. However, due to the small sample statistics, it is not yet possible to conclude, if this morpho-kinematic class of galaxies is systematically faint in the FIR.

\begin{figure}
    \centering
    \includegraphics[width=3.4in]{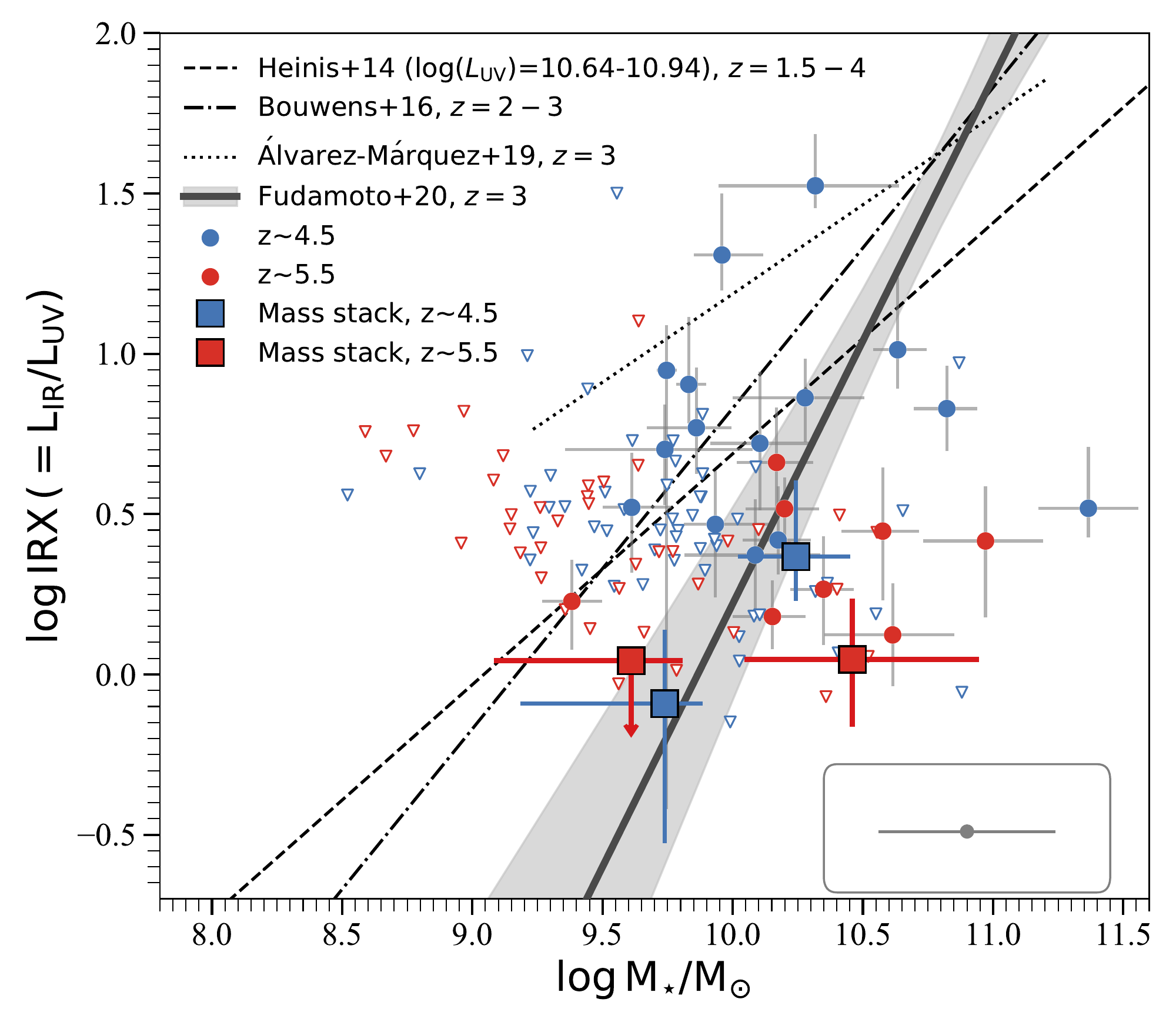}
    \caption{IRX-$M_{\star}$ diagram of our galaxies compared to previously determined relations at different redshifts.
    Blue and red points show individual FIR continuum detections at $4<z<5$ and at $5<z<6$, respectively.
    Open downward triangles show $3\,\sigma$ upper limits of individual IR nondetections.
    Stacks are shown rectangles (blue at $z\sim4.5$, red for $z\sim5.5$).
    The nondetection of the stack is indicated by a downward arrow.
    The inset shows the median uncertainty of individual \Mstar\,estimations ($\rm{log}\,(\Delta M_{\star}/M_{\odot})=\pm0.34$).
    Lines show IRX-\Mstar\,relations derived in previous studies at $z\sim2-4$ \citep{AlvarezMarquez2016,Bouwens2016,Koprowski2018,Fudamoto2019}.
    The gray band shows the $1\,\sigma$ uncertainty of the \citet{Fudamoto2019} relation. With the exception of the high mass bin at $z\sim5.5$, the population average stacked IRX values are consistent with this relation. However, the intrinsic dispersion from galaxy to galaxy at fixed stellar mass is clearly extremely large.
    }
    \label{fig:IRX_Mass_ALPINE}
\end{figure}

\subsection{IRX-$M_{\star}$ relation}
\label{sec:irx-mass}

In this section, we study the IRX-$M_{\star}$ relation of our galaxies by splitting the sample in the two redshift bins $4<z<5$ and $5<z<6$ again, and we compare our results with previously determined IRX-$M_{\star}$ relations at $z\lesssim 4$.

The observed IRX-$M_{\star}$ diagram is shown in Figure \ref{fig:IRX_Mass_ALPINE}.  As is evident, we find a strong redshift dependence of the relation at $z>4$, despite the presence of a very large dispersion within the population.
The individual detections and the stacked IRX values of our sample are generally lower (on average by $\sim0.2\,\rm{dex}$, but reaching up to $\sim1\,\rm{dex}$) than most of the previously determined IRX-$M_{\star}$ relations at lower redshifts $z\sim1.5-4.0$. In particular, in many cases, the $3\,\sigma$ upper limits of individual nondetections at the massive end of our sample are not consistent with previous relations.
While the steep IRX-$M_{\star}$ relation presented in \citet{Fudamoto2019} using $z\sim3$ galaxies is still consistent with our $z\sim4.5$ sample, this is no longer the case at $z\sim5.5$ as the average IRX further decreases by $\sim0.38\,\rm{dex}$ from redshift  $z\sim4.5$ to $z\sim5.5$ (Fig. \ref{fig:IRX_Mass_ALPINE}).

To compare our results in detail with the IRX-$M_{\star}$ relations previously determined at lower redshift, we use relations that are derived from UV selected star forming galaxies at $1.5<z<4$. The IR observations of these studies are based on {\it Herschel} \citep{Heinis2014,AlvarezMarquez2016, AlvarezMarquez2019,Koprowski2018}, or ALMA \citep{Fudamoto2019}.
\citet{Heinis2014} found no evolution of the IRX-$M_{\star}$ relation up to $z\sim4$. However, they do point out that the IRX values decrease at fixed stellar mass as a function of UV luminosity in a way that UV luminous galaxies show systematically lower IRX values than UV fainter ones.
To present a fair comparison, we use the IRX-$M_{\star}$ relation  from \citet{Heinis2014} obtained by stacks of the highest \luv bin ($\rm{log}\,\luv /L_{\odot}=10.64-10.94$), which is  close to the typical \luv\,of our sample (median UV luminosities are $\rm{log}\,\luv /L_{\odot}=10.91$ for $z\sim4.5$ and $\rm{log}\,\luv /L_{\odot}=10.86$ at $z\sim5.5$). Thus, the sample selection bias of our rest-frame UV luminous galaxies has little, if any, effect on the comparison.

The previously derived IRX-$M_{\star}$ relations up to $z\sim4$ are mostly consistent with each other, and thus several studies agree on the nonevolution of the IRX-$M_{\star}$ relation over a wide redshift range.
However, in a recent study, \citet{Fudamoto2019} found a significantly steeper IRX-$M_{\star}$ relation using an unbiased sample of star forming galaxies at $z\sim3$ by exploring the entire public ALMA archive data in COSMOS \citep[A$^3$COSMOS,][]{Liu2019}.
The steeper slope in \citet{Fudamoto2019} could reflect a difference in the FIR SED used in previous studies and/or a potential measurement bias in previous stacking analyses, which relied on low-resolution data (e.g., {\it Herschel}), however the exact reason is not yet clear.

At $z\sim4.5$, in our ALPINE sample, we find that several individual detections are still consistent with previous relations within $3\,\sigma$ uncertainty \citep[e.g.,][]{Heinis2014,Bouwens2016, Fudamoto2019}. However, for individual nondetections at $M_{\star}>10^{10}\,\rm{M_{\odot}}$ where the sensitivity of our observations provide strong constraints, all of our upper limits lie below the previous relations except for 2 galaxies whose upper limits are still consistent with the \citet{Fudamoto2019} relation. 
While the $z\sim4.5$ stacks show significant detections (upper panels of Fig. \ref{fig:massstack}), and are still consistent with the steep IRX-$M_{\star}$ relation from \citet{Fudamoto2019}, the $z\sim4.5$ stacks show much lower IRX than the previously determined relations at $z\sim3$ including the most UV luminous bin from \citet{Heinis2014}.
Comparing our stacks with the IRX-$M_{\star}$ relation of \citet{Bouwens2016}, the IRX of $z\sim4.5$ galaxies, on average, are $\sim0.63\,\rm{dex}$ lower at a fixed stellar mass.

Based on these considerations, we conclude the IRX-$M_{\star}$ relation of our sample at $z\sim4.5$ is consistent with that of \citet{Fudamoto2019}, and deviate from the other previously found relations. This suggests that the IRX-$M_{\star}$ relation of main-sequence galaxies at $4<z<5$ either rapidly evolves and shows $0.6\,\rm{dex}$ lower IRX at a fixed stellar mass than the $z\lesssim4$ relations, or is consistent with the steep IRX-$M_{\star}$ relation of \citet{Fudamoto2019} found at $z\sim3$.

At $z\sim5.5$ this situation changes rapidly as almost all the individual detections lie below the IRX-$M_{\star}$ relations at $z\lesssim4$, and, at the massive end of our $z\sim5.5$ sample (i.e., $M_{\star}>10^{10}\,\rm{M_{\star}}$), all the individual $3\,\sigma$ upper limits are below the $z\lesssim4$ IRX-$M_{\star}$ relations, except for one upper limit that is still consistent with \citet{Fudamoto2019}.
Stacking results emphasise the discrepancies between our $z\sim5.5$ sample and the $z\lesssim4$ IRX-$M_{\star}$ relations. 
While the stack of the lower mass bin ($M_{\star}=10^{9}-10^{10}\,\rm{M_{\star}}$) does not show a significant detection (lower left panel of Fig. \ref{fig:massstack}), its $3\,\sigma$ upper limit lies $\sim0.4\,\rm{dex}$ below the \citet{Heinis2014} and the \citet{Bouwens2016} IRX-$M_{\star}$ relations.
The stack of higher mass bin ($M_{\star}=10^{10}-10^{11}\,\rm{M_{\star}}$) shows a detection (lower right panel of Fig. \ref{fig:massstack}), however its IRX is $\gtrsim1\,\rm{dex}$ below all the previously estimated IRX-$M_{\star}$ relations, suggesting that previously known IRX-$M_{\star}$ relations could over-predict the IR luminosities of $z>5$ star forming galaxies by $\sim1\,\rm{dex}$.

The overall comparisons above demonstrate that the IRX-$M_{\star}$ relations from our observations start to become inconsistent with the previously determined IRX-$M_{\star}$ relations from $z<4$ (except with steeper relation as in \citet{Fudamoto2019}), and is no longer consistent with all the previously derived IRX-$M_{\star}$ relations by $z\sim5.5$.
The rapid decrease of the IRX from $z\lesssim4$ to $z>4.5$ in massive galaxies suggests a rapid evolution of dust attenuation properties of star formation in the high-redshift Universe, consistent with our conclusions from the IRX-$\beta$ diagram.

\begin{figure}[tbp]
    \centering
    \includegraphics[width=3.6in]{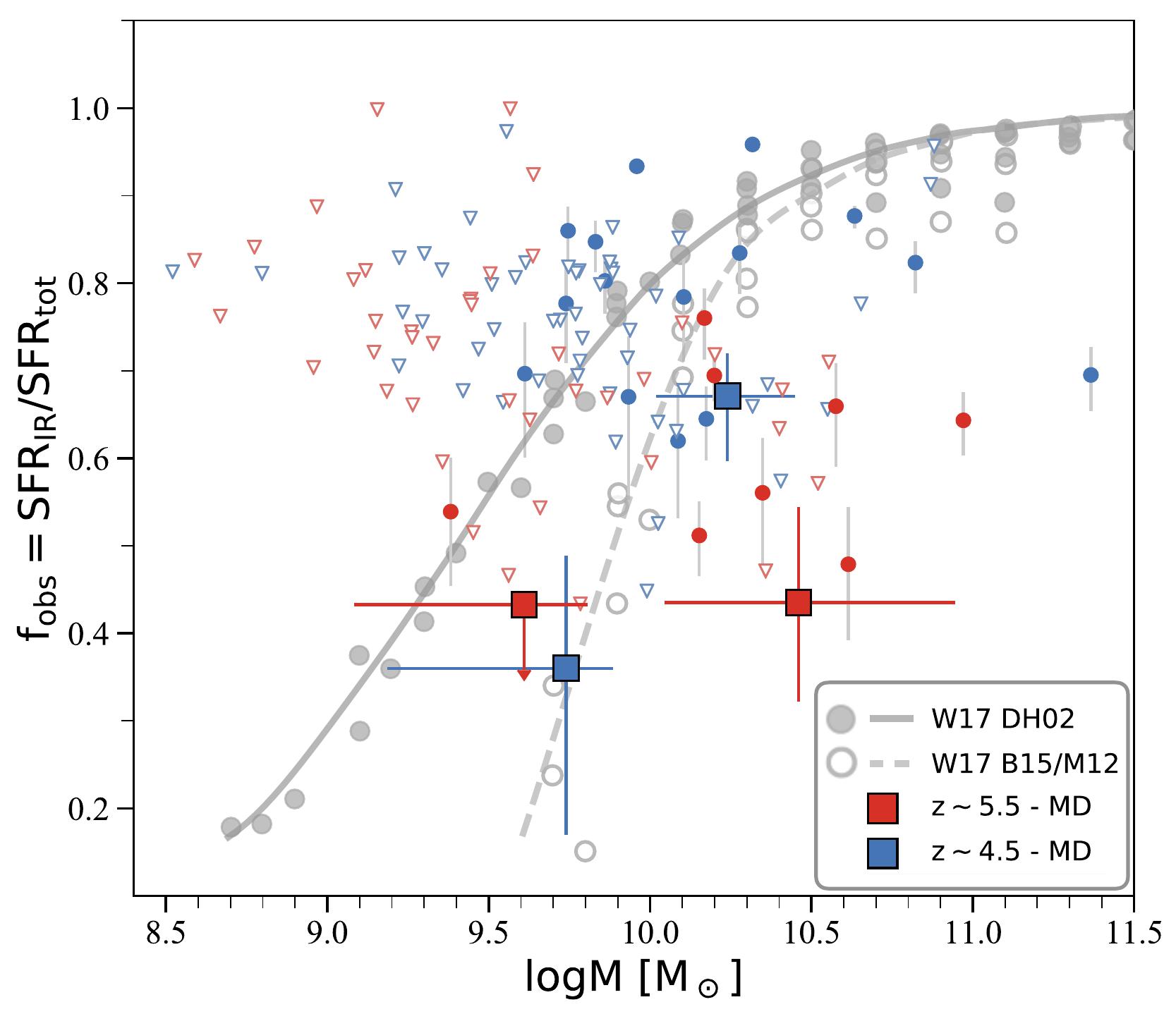}
    \caption{
    Obscured fraction of star formation as a function of stellar mass ($f_{\rm obs}-M_{\star}$ relation) of our UV selected sample.
    Blue and red points show individual FIR continuum detections at $4<z<5$ and at $5<z<6$, respectively.
    Triangles show $3\sigma$ upper limits for IR nondetections.
    Stacks are shown by blue (at $z\sim4.5$) and by red (at $z\sim5.5$) rectangles.
    The nondetection of the stack is indicated by a downward arrow.
    Gray points and lines show the observed relation at redshifts between $z=0$ and $z\sim2.5$ \citep{Whitaker2017}.
    The solid line shows the constant $f_{\rm obs}$-$M_{\star}$ relation of $z\sim2.5$ to $z\sim0$ using a template from \citet{Dale2002}, and the dashed line shows the same using \citet{Bethermin2015} or \citet{Magdis2012} templates.
    At $M_{\star}<10^{10}\,\rm{M_{\odot}}$, our $z\sim4.5$ stacks are potentially consistent with the $f_{\rm obs}$ at $z\sim2.5$ to $z\sim0$ using \citet{Bethermin2015} or \citet{Magdis2012} templates.
    However, at $M_{\star}>10^{10}\,\rm{M_{\odot}}$, our stacks show decreasing $f_{\rm obs}$ from $z\sim2.5$ to $z\sim5.5$, suggesting a rapid evolution of dust obscured star formation activity in main-sequence galaxies at $z>4$. However, we caution that UV-selected samples at $z>4$ may be incomplete at very high masses ($M_{\star}>10^{10.5}\,\rm{M_{\odot}}$) given the absence of deep rest-frame optical imaging, as as been shown recently by the detection of a significant population of UV-undetected, massive galaxies \citep[e.g.,][]{Wang19,AlcaldePampliega19}. 
    }
    \label{fig:Fobs}
\end{figure}

\subsection{The obscured fraction of star formation}
\label{sec:fobs}

Having estimated both the UV-based SFR and the infrared-based SFRs for our galaxies, we can estimate the obscured fraction of star formation occurring at $z>4$. This obscured fraction, $f_{\rm obs}$, represents the fraction of star formation activity observable from IR continuum emission (i.e., dust obscured star formation activity) relative to its total amount, and is defined as 
$f_{\rm obs}={\rm SFR_{IR}/SFR_{tot}}$,
where $\rm{SFR_{IR}}$ is the star formation rate observed from IR (equation \ref{eqn:sfr_ir}) and $\rm{SFR_{tot}}$ is the total star formation rate obtained by adding UV and IR based SFR estimates (i.e., Equation \ref{eqn:sfr_uv} + Equation \ref{eqn:sfr_ir}).


Using Spitzer MIPS $24\,\rm{\mu m}$ observations of a mass complete sample at ${\rm log}\,M_\ast/M_\odot\gtrsim9$, \citet{Whitaker2017} show that $f_{\rm obs}$ is highly mass dependent, with more than 80\% of star formation being obscured in massive galaxies with ${\rm log}\,M_\ast/M_\odot\gtrsim10$. Remarkably, \citet{Whitaker2017} find that this $f_{\rm obs}-M_{\star}$ relation remains constant over the full redshift range $z\sim2.5$ to $z\sim0$.
Since $f_{\rm obs}$ is directly related to the IRX, the nonevolution of the $f_{\rm obs} - M_\star$ relation can be considered a product of the nonevolution of the IRX-$M_\star$ relation observed at $z\sim1.5-3$ \citep[e.g.,][]{Heinis2014}.
In the same way, our finding of an evolving IRX-$M_\star$ relation at $z>4.5$ (Section \ref{sec:irx-mass}) thus implies an evolution of the obscured fraction of star formation at $z>4$.

Figure \ref{fig:Fobs} presents the $f_{\rm obs} - M_\star$ diagram of our galaxy sample.
While individual detections and upper limits generally lie between $f_\mathrm{obs}=60-90\%$, given our ALMA sensitivity limits, the population average stacked values are significantly lower for 3 out of our 4 stacks.  In particular, at $M_{\star}>10^{10} \, \rm{M_{\odot}}$, our stacks show lower values than the $f_{\rm obs}$-$M_{\star}$ relation of \citet{Whitaker2017}. While $z=0-2.5$ galaxies reach $f_{\rm obs}\gtrsim80\,\%$ at these masses, we find $f_{\rm obs}=0.67^{+0.05}_{-0.07}$ at $z\sim4.5$ and only $f_{\rm obs}=0.44^{+0.11}_{-0.11}$ at $z\sim5.5$.
Remarkably, in this high-mass bin at $z\sim5.5$, even all the individual detections and upper limits lie significantly below the lower redshift $f_{\rm obs}$-$M_{\star}$ relation.

At $M_{\star}<10^{10} \, \rm{M_{\odot}}$, \citet{Whitaker2017} discussed that the estimated \lir\,systematically changes depending on the FIR SED templates used. Using their default template from \citet{Dale2002}, they find a significantly shallower trend with mass compared to a template set based on \citet{Bethermin2015} and \citet{Magdis2012}, which incorporate an evolution of dust temperature as a function of redshift. Figure \ref{fig:Fobs} shows both these trends. While the mass trend changes, \citet{Whitaker2017} show that the nonevolution of the $f_{\rm obs}$-$M_{\star}$ relation up to $z\sim2.5$ is conserved, independent of the template choice.

Our stacked data points at $M_{\star}<10^{10}\,\rm{M_{\odot}}$ result in a mean of $f_{\rm obs}=0.36^{+0.13}_{-0.19}$ at $z\sim4.5$ and a 3$\sigma$ upper limit of $f_{\rm obs}<0.43$ at $z\sim5.5$. As seen in Figure \ref{fig:Fobs}, these numbers are consistent  with the lower redshift $f_{\rm obs}$-$M_{\star}$ relation of \citet{Whitaker2017}
using the \citet{Bethermin2015} and \citet{Magdis2012} templates.
However, the individual detection and the stacked data points at $M_{\star}>10^{10}\,\rm{M_{\odot}}$ show rapid decrease of the obscured fractions at $z>4.5$ for both template sets.
This may thus indicate a different redshift evolution for low and high mass galaxies. It is also clear that the obscured fraction of star formation varies significantly from galaxy to galaxy, as is evident from the many individual continuum detections of low-mass galaxies at $z\sim4.5$, which imply $f_{\rm obs}>0.7$, while the stacked median value is only $f_{\rm obs}=0.36^{+0.13}_{-0.19}$.




One likely caveat of our analysis is that our sample is not perfectly mass complete. While ALPINE galaxies are selected to lie on the main-sequence, they were all required to have spectroscopic redshift measurements, which in most cases are based on rest-frame UV spectra. This could potentially bias our sample to somewhat more UV-luminous, less obscured systems. However, \citet{Faisst2019} show that the ALPINE sample only exhibits a weak bias toward bluer UV continuum slopes compared to a mass-selected parent sample with photometric redshifts $z_\mathrm{phot}=4-6$. Additionally, the UV selection should affect the $z\sim4.5$ or $z\sim5.5$ galaxies in a similar way. Simple tests comparing the ALPINE sample with a mass-matched COSMOS parent sample indeed do not find systematic differences in IRX measurements as a function of UV slope.

Nevertheless,
it is clear that extremely obscured, dusty sources, which still lie on the main sequence, would be missing from our sample. The existence of such a population of very massive, dusty galaxies at $z>3$, which can remain undetected at rest-frame UV wavelengths, has recently been suggested in the literature based on ALMA or {\it Spitzer} detections \citep[e.g.,][]{Franco18,Wang19, Williams19,Casey19,Gruppioni2020}. The exact contribution to the total SFRD of such sources is still uncertain, given the currently small sample sizes and the difficulty of measuring their exact redshifts and stellar masses. However, it is clear that such UV-faint galaxies have the potential to dominate the SFRD at the massive end of the $z>3$ galaxy population, at $M_{\star}\gtrsim 10^{10.5}\,\rm{M_{\odot}}$. Solving this question will likely have to await the advent of the James Webb Space Telescope, which will provide much deeper rest-frame optical observations.

Keeping this potential sample bias at very high masses in mind, we can nevertheless conclude that our \textit{UV selected}, main-sequence galaxies at $M_{\star}>10^{10} \,\rm{M_{\odot}}$ show a much lower obscured fraction than galaxies at $z\lesssim3$. This implies a very rapid build-up of dust in such massive galaxies in the early universe, as the obscured fraction increases from $\sim45\%$ to $\gtrsim80\%$ between $z\sim5.5$ to $z\sim2.5$. In contrast, at lower masses, our stacked measurements are consistent with the constant lower redshift $f_{\rm obs}$-$M_{\star}$ relation of \citet{Whitaker2017} with \citet{Bethermin2015} SED templates, which imply $f_{\rm obs}\lesssim45\%$ at $M_{\star}<10^{10} \,\rm{M_{\odot}}$.




\section{Summary and conclusions}
\label{sec:conclusion}

We have examined the IRX-$\beta$ relation, the IRX-$M_{\star}$ relation, and the obscured fraction of star formation as a function of stellar mass ($f_{\rm obs}$-$M_{\star}$ relation) of UV-selected main-sequence star forming galaxies at $z\sim4.4-5.8$ using $\lambda_{\rm rest}=158\,\rm{\mu m}$ continuum observations of 118 galaxies from the ALMA large program, ALPINE.
The sample has secure spectroscopic redshifts, but is nevertheless representative of star forming galaxies, that is to say the distribution of SFRs and $M_{\star}$ are consistent with normal main-sequence galaxies in the observed redshift range: $\rm{log\,(M_{\star}/M_{\odot}) \sim 8.5 - 11.5}$ and $\rm{log\,(SFR/M_{\odot}\,yr^{-1}}) \sim 0.5 - 2.5$ \citep[see][]{Faisst2019}.
From individual FIR measurements and stacks of both detections and nondetections we found:
\vspace{7pt}

(i) The IRX-$\beta$ relation of our sample (Fig \ref{fig:IRX_ALPINE}) is generally located below the Meurer relation \citep{Meurer1999} with average IRX values $>0.5$dex lower for galaxies redder than $\beta_{UV}>-1.5$, as hinted at by earlier studies \citep[e.g.,][]{Capak2015,Barisic2017}.
Individual measurements of the UV spectral slope $\beta$ suggest that the intrinsic (or dust free) UV spectral slope $\beta_0$ is smaller than the locally estimated value of $\beta_0=-2.23$ \citep{Meurer1999}, and more consistent with a bluer $\beta_0$, as expected for younger, more metal poor galaxies \citep[e.g., $\beta_0=-2.62$,][]{Reddy2018}.
We fit for the slope of the UV attenuation curve, finding it to be significantly steeper than that of local star forming galaxies, and similar to, or even steeper than that of the SMC (treating the SMC extinction curve as an attenuation curve; see Section \ref{sec:IRX-beta}).
\vspace{7pt}

(ii) Most of the IRX-$M_{\star}$ relations previously reported in the literature at $z\lesssim3$ are not consistent with the population average relation of our galaxies at $z=4.4-5.8$ (Fig \ref{fig:IRX_Mass_ALPINE}).
At $z\sim4.5$, our sample is still consistent with the steep IRX-$M_{\star}$ relation found by \citet{Fudamoto2019} for $z\sim3$ galaxies, even though individually detected galaxies show a very large scatter around the mean relation. However, at $z\sim5.5$, all the previously found IRX-$M_{\star}$ relations over-predict the IR luminosities by $\sim1\,{\rm dex}$, in particular for galaxies at the high-mass end of our sample, $\rm{log\,(M_{\star}/M_{\odot}) >10.0}$.
\vspace{7pt}

(iii) The fraction of dust-obscured star formation among our UV-selected sample shows a decrease from $f_{\rm obs} \sim65\,\%$ at $z\sim4.5$ to only $\sim45\,\%$ at $z\sim5.5$ in 
massive ($M_{\star} \sim (1-3) \times 10^{10}\,\rm{M_{\odot}}$) galaxies with a potentially large scatter. This average obscured fraction is significantly lower than the $\gtrsim80\%$ found for lower redshift galaxies, and provides direct evidence that the obscured fraction of star forming galaxies rapidly evolves in the early universe.
\vspace{7pt}

Taken together, our results indicate that the dust attenuation properties evolve rapidly between $z\sim6$, at the end of cosmic reionization, to $z\sim2-3$, around the peak of cosmic star formation. It is possible that we are seeing a transition to supernovae (SNe) driven dust production at $z\gtrsim3$, which is predicted theoretically given the limited time available for dust production in lower mass stars \citep[e.g.,][]{Todini01,Nozawa03,Schneider04}. Under certain conditions, such SNe dust might also be consistent with the steeper dust curve we find for our $z\sim4-6$ galaxy sample compared to the M99-like attenuation inferred at $z<3$ \citep[e.g.,][]{Maiolino2004,Hirashita05,Stratta2007,Gallerani10}. We refer to a future paper to analyze this possibility in detail.

Importantly, our results indicate that the previous dust attenuation corrections calibrated at $z<4$ will over-predict the \lir\,of higher-redshift UV selected, star forming galaxies by up to $\sim1\,{\rm dex}$, especially for red and relatively massive galaxies.
Future multiwavelength observations of UV-red galaxies (i.e., with $\beta > -1.5$) will be crucial to further constrain the dust attenuation properties of galaxies at $5<z<6$ and to obtain a complete census of the total SFR density in the early universe. This is one of the key questions that can be addressed in the near future by exploiting the synergy between the FIR observations from ALMA and the rest-frame optical data coming from the James Webb Space Telescope.



\vspace{7pt}


\begin{acknowledgements}
This paper is dedicated to the memory of Olivier Le F\`evre, PI of the ALPINE survey. We thank the anonymous referee for insightful comments that improved the quality of this manuscript.
This paper makes use of the following ALMA data:
 ADS/JAO.ALMA\#2017.1.00428.L. ALMA is a partnership of ESO
(representing its member states), NSF (USA) and NINS (Japan), together
with NRC (Canada) and NSC and ASIAA (Taiwan) and KASI (Republic of
Korea), in cooperation with the Republic of Chile. 
The Joint ALMA Observatory is operated by ESO, AUI/NRAO and NAOJ.
Based on data products from observations made with ESO Telescopes at the La Silla Paranal Observatory under ESO programme ID 179.A-2005 and on data products produced by TERAPIX and the Cambridge Astronomy Survey Unit on behalf of the UltraVISTA consortium. This work was supported by the Swiss
National Science Foundation through the SNSF Professorship grant
157567 'Galaxy Build-up at Cosmic Dawn'.
G.C.J. acknowledges ERC Advanced Grant 695671 ``QUENCH'' and support by the Science and Technology Facilities Council (STFC).
D.R. acknowledges support from the National Science Foundation under grant number AST-1614213 and from the Alexander von Humboldt Foundation through a Humboldt Research Fellowship for Experienced Researchers.
A.C., F.P. and M.T. acknowledge the grant MIUR PRIN2017.
LV acknowledges funding from the European Union’s Horizon 2020 research and innovation program under the Marie Sklodowska-Curie Grant agreement No. 746119.
S.F. is supported by and the Cosmic Dawn Center of Excellence funded by the Danish National Research Foundation under then grant No. 140.
R.A. acknowledges support from FONDECYT Regular Grant 1202007.
\end{acknowledgements}

%
\bibliographystyle{aa} 
\bibliography{./ref.bib} 

\begin{thebibliography}{91}
\expandafter\ifx\csname natexlab\endcsname\relax\def\natexlab#1{#1}\fi

\bibitem[{{Alavi} {et~al.}(2014){Alavi}, {Siana}, {Richard}, {Stark},
  {Scarlata}, {Teplitz}, {Freeman}, {Dominguez}, {Rafelski}, {Robertson}, \&
  {Kewley}}]{Alavi2014}
{Alavi}, A., {Siana}, B., {Richard}, J., {et~al.} 2014, \apj, 780, 143

\bibitem[{{Alcalde Pampliega} {et~al.}(2019){Alcalde Pampliega},
  {P{\'e}rez-Gonz{\'a}lez}, {Barro}, {Dom{\'\i}nguez S{\'a}nchez},
  {Eliche-Moral}, {Cardiel}, {Hern{\'a}n-Caballero}, {Rodriguez-Mu{\~n}oz},
  {S{\'a}nchez Bl{\'a}zquez}, \& {Esquej}}]{AlcaldePampliega19}
{Alcalde Pampliega}, B., {P{\'e}rez-Gonz{\'a}lez}, P.~G., {Barro}, G., {et~al.}
  2019, \apj, 876, 135

\bibitem[{{{\'A}lvarez-M{\'a}rquez} {et~al.}(2019){{\'A}lvarez-M{\'a}rquez},
  {Burgarella}, {Buat}, {Ilbert}, \&
  {P{\'e}rez-Gonz{\'a}lez}}]{AlvarezMarquez2019}
{{\'A}lvarez-M{\'a}rquez}, J., {Burgarella}, D., {Buat}, V., {Ilbert}, O., \&
  {P{\'e}rez-Gonz{\'a}lez}, P.~G. 2019, \aap, 630, A153

\bibitem[{{{\'A}lvarez-M{\'a}rquez} {et~al.}(2016){{\'A}lvarez-M{\'a}rquez},
  {Burgarella}, {Heinis}, {Buat}, {Lo Faro}, {B{\'e}thermin},
  {L{\'o}pez-Fort{\'{\i}}n}, {Cooray}, {Farrah}, {Hurley}, {Ibar}, {Ilbert},
  {Koekemoer}, {Lemaux}, {P{\'e}rez-Fournon}, {Rodighiero}, {Salvato}, {Scott},
  {Taniguchi}, {Vieira}, \& {Wang}}]{AlvarezMarquez2016}
{{\'A}lvarez-M{\'a}rquez}, J., {Burgarella}, D., {Heinis}, S., {et~al.} 2016,
  \aap, 587, A122

\bibitem[{{Aretxaga} {et~al.}(2011){Aretxaga}, {Wilson}, {Aguilar}, {Alberts},
  {Scott}, {Scoville}, {Yun}, {Austermann}, {Downes}, {Ezawa}, {Hatsukade},
  {Hughes}, {Kawabe}, {Kohno}, {Oshima}, {Perera}, {Tamura}, \&
  {Zeballos}}]{Aretxaga2011}
{Aretxaga}, I., {Wilson}, G.~W., {Aguilar}, E., {et~al.} 2011, \mnras, 415,
  3831

\bibitem[{{Arnouts} {et~al.}(1999){Arnouts}, {Cristiani}, {Moscardini},
  {Matarrese}, {Lucchin}, {Fontana}, \& {Giallongo}}]{Arnouts1999}
{Arnouts}, S., {Cristiani}, S., {Moscardini}, L., {et~al.} 1999, \mnras, 310,
  540

\bibitem[{{Barisic} {et~al.}(2017){Barisic}, {Faisst}, {Capak}, {Pavesi},
  {Riechers}, {Scoville}, {Cooke}, {Kartaltepe}, {Casey}, \&
  {Smolcic}}]{Barisic2017}
{Barisic}, I., {Faisst}, A.~L., {Capak}, P.~L., {et~al.} 2017, \apj, 845, 41

\bibitem[{{B{\'e}thermin} {et~al.}(2015){B{\'e}thermin}, {Daddi}, {Magdis},
  {Lagos}, {Sargent}, {Albrecht}, {Aussel}, {Bertoldi}, {Buat}, {Galametz},
  {Heinis}, {Ilbert}, {Karim}, {Koekemoer}, {Lacey}, {Le Floc'h}, {Navarrete},
  {Pannella}, {Schreiber}, {Smol{\v{c}}i{\'c}}, {Symeonidis}, \&
  {Viero}}]{Bethermin2015}
{B{\'e}thermin}, M., {Daddi}, E., {Magdis}, G., {et~al.} 2015, \aap, 573, A113

\bibitem[{{Bethermin} {et~al.}(2020){Bethermin}, {Fudamoto}, {Ginolfi},
  {Loiacono}, {Khusanova}, {Capak}, {Cassata}, {Faisst}, {Le Fevre},
  {Schaerer}, {Silverman}, {Yan}, {Amorin}, {Bardelli}, {Boquien}, {Cimatti},
  {Davidzon}, {Dessauges-Zavadsky}, {Fujimoto}, {Gruppioni}, {Hathi}, {Ibar},
  {Jones}, {Koekemoer}, {Lagache}, {Lemaux}, {Oesch}, {Pozzi}, {Riechers},
  {Talia}, {Toft}, {Vallini}, {Vergani}, {Zamorani}, \&
  {Zucca}}]{Bethermin2019}
{Bethermin}, M., {Fudamoto}, Y., {Ginolfi}, M., {et~al.} 2020, arXiv e-prints,
  arXiv:2002.00962

\bibitem[{{B{\'e}thermin} {et~al.}(2017){B{\'e}thermin}, {Wu}, {Lagache},
  {Davidzon}, {Ponthieu}, {Cousin}, {Wang}, {Dor{\'e}}, {Daddi}, \&
  {Lapi}}]{Bethermin2017}
{B{\'e}thermin}, M., {Wu}, H.-Y., {Lagache}, G., {et~al.} 2017, \aap, 607, A89

\bibitem[{{Bourne} {et~al.}(2017){Bourne}, {Dunlop}, {Merlin}, {Parsa},
  {Schreiber}, {Castellano}, {Conselice}, {Coppin}, {Farrah}, {Fontana},
  {Geach}, {Halpern}, {Knudsen}, {Micha{\l}owski}, {Mortlock}, {Santini},
  {Scott}, {Shu}, {Simpson}, {Simpson}, {Smith}, \& {van der
  Werf}}]{Bourne2017}
{Bourne}, N., {Dunlop}, J.~S., {Merlin}, E., {et~al.} 2017, \mnras, 467, 1360

\bibitem[{{Bouwens} {et~al.}(2016){Bouwens}, {Aravena}, {Decarli}, {Walter},
  {da Cunha}, {Labb{\'e}}, {Bauer}, {Bertoldi}, {Carilli}, {Chapman}, {Daddi},
  {Hodge}, {Ivison}, {Karim}, {Le Fevre}, {Magnelli}, {Ota}, {Riechers},
  {Smail}, {van der Werf}, {Weiss}, {Cox}, {Elbaz}, {Gonzalez-Lopez},
  {Infante}, {Oesch}, {Wagg}, \& {Wilkins}}]{Bouwens2016}
{Bouwens}, R.~J., {Aravena}, M., {Decarli}, R., {et~al.} 2016, \apj, 833, 72

\bibitem[{{Bouwens} {et~al.}(2015){Bouwens}, {Illingworth}, {Oesch}, {Trenti},
  {Labb{\'e}}, {Bradley}, {Carollo}, {van Dokkum}, {Gonzalez}, {Holwerda},
  {Franx}, {Spitler}, {Smit}, \& {Magee}}]{Bouwens2015}
{Bouwens}, R.~J., {Illingworth}, G.~D., {Oesch}, P.~A., {et~al.} 2015, \apj,
  803, 34

\bibitem[{{Bouwens} {et~al.}(2019){Bouwens}, {Stefanon}, {Oesch},
  {Illingworth}, {Nanayakkara}, {Roberts-Borsani}, {Labb{\'e}}, \&
  {Smit}}]{Bouwens2019}
{Bouwens}, R.~J., {Stefanon}, M., {Oesch}, P.~A., {et~al.} 2019, \apj, 880, 25

\bibitem[{{Bowler} {et~al.}(2018){Bowler}, {Bourne}, {Dunlop}, {McLure}, \&
  {McLeod}}]{Bowler18}
{Bowler}, R.~A.~A., {Bourne}, N., {Dunlop}, J.~S., {McLure}, R.~J., \&
  {McLeod}, D.~J. 2018, \mnras, 481, 1631

\bibitem[{{Calzetti} {et~al.}(2000){Calzetti}, {Armus}, {Bohlin}, {Kinney},
  {Koornneef}, \& {Storchi-Bergmann}}]{Calzetti2000}
{Calzetti}, D., {Armus}, L., {Bohlin}, R.~C., {et~al.} 2000, \apj, 533, 682

\bibitem[{{Capak} {et~al.}(2015){Capak}, {Carilli}, {Jones}, {Casey},
  {Riechers}, {Sheth}, {Carollo}, {Ilbert}, {Karim}, \& {Lefevre}}]{Capak2015}
{Capak}, P.~L., {Carilli}, C., {Jones}, G., {et~al.} 2015, \nat, 522, 455

\bibitem[{{Casey} {et~al.}(2013){Casey}, {Chen}, {Cowie}, {Barger}, {Capak},
  {Ilbert}, {Koss}, {Lee}, {Le Floc'h}, {Sand ers}, \& {Williams}}]{Casey2013}
{Casey}, C.~M., {Chen}, C.-C., {Cowie}, L.~L., {et~al.} 2013, \mnras, 436, 1919

\bibitem[{{Casey} {et~al.}(2014){Casey}, {Scoville}, {Sanders}, {Lee},
  {Cooray}, {Finkelstein}, {Capak}, {Conley}, {De Zotti}, {Farrah}, {Fu}, {Le
  Floc'h}, {Ilbert}, {Ivison}, \& {Takeuchi}}]{Casey2014}
{Casey}, C.~M., {Scoville}, N.~Z., {Sanders}, D.~B., {et~al.} 2014, \apj, 796,
  95

\bibitem[{{Casey} {et~al.}(2019){Casey}, {Zavala}, {Aravena}, {B{\'e}thermin},
  {Caputi}, {Champagne}, {Clements}, {da Cunha}, {Drew}, {Finkelstein},
  {Hayward}, {Kartaltepe}, {Knudsen}, {Koekemoer}, {Magdis}, {Man}, {Manning},
  {Scoville}, {Sheth}, {Spilker}, {Staguhn}, {Talia}, {Taniguchi}, {Toft},
  {Treister}, \& {Yun}}]{Casey19}
{Casey}, C.~M., {Zavala}, J.~A., {Aravena}, M., {et~al.} 2019, \apj, 887, 55

\bibitem[{{Castellano} {et~al.}(2014){Castellano}, {Sommariva}, {Fontana},
  {Pentericci}, {Santini}, {Grazian}, {Amorin}, {Donley}, {Dunlop}, {Ferguson},
  {Fiore}, {Galametz}, {Giallongo}, {Guo}, {Huang}, {Koekemoer}, {Maiolino},
  {McLure}, {Paris}, {Schaerer}, {Troncoso}, \& {Vanzella}}]{Castellano2014}
{Castellano}, M., {Sommariva}, V., {Fontana}, A., {et~al.} 2014, \aap, 566, A19

\bibitem[{{Chabrier}(2003)}]{Chabrier2003}
{Chabrier}, G. 2003, \pasp, 115, 763

\bibitem[{{Condon}(1997)}]{Condon1997}
{Condon}, J.~J. 1997, \pasp, 109, 166

\bibitem[{{Dale} \& {Helou}(2002)}]{Dale2002}
{Dale}, D.~A. \& {Helou}, G. 2002, \apj, 576, 159

\bibitem[{{Davidzon} {et~al.}(2017){Davidzon}, {Ilbert}, {Laigle}, {Coupon},
  {McCracken}, {Delvecchio}, {Masters}, {Capak}, {Hsieh}, {Le F{\`e}vre},
  {Tresse}, {Bethermin}, {Chang}, {Faisst}, {Le Floc'h}, {Steinhardt}, {Toft},
  {Aussel}, {Dubois}, {Hasinger}, {Salvato}, {Sanders}, {Scoville}, \&
  {Silverman}}]{Davidzon2017}
{Davidzon}, I., {Ilbert}, O., {Laigle}, C., {et~al.} 2017, \aap, 605, A70

\bibitem[{{Eldridge} \& {Stanway}(2012)}]{Eldridge2012}
{Eldridge}, J.~J. \& {Stanway}, E.~R. 2012, \mnras, 419, 479

\bibitem[{{Faisst} {et~al.}(2017){Faisst}, {Capak}, {Yan}, {Pavesi},
  {Riechers}, {Bari{\v{s}}i{\'c}}, {Cooke}, {Kartaltepe}, \&
  {Masters}}]{Faisst2017}
{Faisst}, A.~L., {Capak}, P.~L., {Yan}, L., {et~al.} 2017, \apj, 847, 21

\bibitem[{{Faisst} {et~al.}(2019){Faisst}, {Schaerer}, {Lemaux}, {Oesch},
  {Fudamoto}, {Cassata}, {Bethermin}, {Capak}, {Le Fevre}, {Silverman}, {Yan},
  {Ginolfi}, {Koekemoer}, {Morselli}, {Amorin}, {Bardelli}, {Boquien},
  {Brammer}, {Cimatti}, {Dessauges-Zavadsky}, {Fujimoto}, {Gruppioni}, {Hathi},
  {Hemmati}, {Jones}, {Khusanova}, {Loiacono}, {Pozzi}, {Talia}, {Tasca},
  {Riechers}, {Rodighiero}, {Romano}, {Scoville}, {Toft}, {Vallini}, {Vergani},
  {Zamorani}, \& {Zucca}}]{Faisst2019}
{Faisst}, A.~L., {Schaerer}, D., {Lemaux}, B.~C., {et~al.} 2019, arXiv
  e-prints, arXiv:1912.01621

\bibitem[{{Ferrara} {et~al.}(2017){Ferrara}, {Hirashita}, {Ouchi}, \&
  {Fujimoto}}]{Ferrara2017}
{Ferrara}, A., {Hirashita}, H., {Ouchi}, M., \& {Fujimoto}, S. 2017, \mnras,
  471, 5018

\bibitem[{{Franco} {et~al.}(2018){Franco}, {Elbaz}, {B{\'e}thermin},
  {Magnelli}, {Schreiber}, {Ciesla}, {Dickinson}, {Nagar}, {Silverman},
  {Daddi}, {Alexander}, {Wang}, {Pannella}, {Le Floc'h}, {Pope}, {Giavalisco},
  {Maury}, {Bournaud}, {Chary}, {Demarco}, {Ferguson}, {Finkelstein}, {Inami},
  {Iono}, {Juneau}, {Lagache}, {Leiton}, {Lin}, {Magdis}, {Messias},
  {Motohara}, {Mullaney}, {Okumura}, {Papovich}, {Pforr}, {Rujopakarn},
  {Sargent}, {Shu}, \& {Zhou}}]{Franco18}
{Franco}, M., {Elbaz}, D., {B{\'e}thermin}, M., {et~al.} 2018, \aap, 620, A152

\bibitem[{{Fudamoto} {et~al.}(2020){Fudamoto}, {Oesch}, {Magnelli},
  {Schinnerer}, {Liu}, {Lang}, {Jim{\'e}nez-Andrade}, {Groves}, {Leslie}, \&
  {Sargent}}]{Fudamoto2019}
{Fudamoto}, Y., {Oesch}, P.~A., {Magnelli}, B., {et~al.} 2020, \mnras, 491,
  4724

\bibitem[{{Fudamoto} {et~al.}(2017){Fudamoto}, {Oesch}, {Schinnerer}, {Groves},
  {Karim}, {Magnelli}, {Sargent}, {Cassata}, {Lang}, \& {Liu}}]{Fudamoto2017}
{Fudamoto}, Y., {Oesch}, P.~A., {Schinnerer}, E., {et~al.} 2017, \mnras, 472,
  483

\bibitem[{{Fujimoto} {et~al.}(2020){Fujimoto}, {Silverman}, {Bethermin},
  {Ginolfi}, {Jones}, {Le F{\`e}vre}, {Dessauges-Zavadsky}, {Rujopakarn},
  {Faisst}, {Fudamoto}, {Cassata}, {Morselli}, {Schaerer}, {Capak}, {Yan},
  {Vallini}, {Toft}, {Loiacono}, {Zamorani}, {Talia}, {Narayanan}, {Hathi},
  {Lemaux}, {Boquien}, {Amorin}, {Ibar}, {Koekemoer},
  {M{\'e}ndez-Hern{\'a}ndez}, {Bardelli}, {Vergani}, {Zucca}, {Romano}, \&
  {Cimatti}}]{Fujimoto2020}
{Fujimoto}, S., {Silverman}, J.~D., {Bethermin}, M., {et~al.} 2020, arXiv
  e-prints, arXiv:2003.00013

\bibitem[{{Gallerani} {et~al.}(2010){Gallerani}, {Maiolino}, {Juarez}, {Nagao},
  {Marconi}, {Bianchi}, {Schneider}, {Mannucci}, {Oliva}, {Willott}, {Jiang},
  \& {Fan}}]{Gallerani10}
{Gallerani}, S., {Maiolino}, R., {Juarez}, Y., {et~al.} 2010, \aap, 523, A85

\bibitem[{{Ginolfi} {et~al.}(2020){Ginolfi}, {Jones}, {B{\'e}thermin},
  {Fudamoto}, {Loiacono}, {Fujimoto}, {Le F{\'e}vre}, {Faisst}, {Schaerer},
  {Cassata}, {Silverman}, {Yan}, {Capak}, {Bardelli}, {Boquien}, {Carraro},
  {Dessauges-Zavadsky}, {Giavalisco}, {Gruppioni}, {Ibar}, {Khusanova},
  {Lemaux}, {Maiolino}, {Narayanan}, {Oesch}, {Pozzi}, {Rodighiero}, {Talia},
  {Toft}, {Vallini}, {Vergani}, \& {Zamorani}}]{Ginolfi2019}
{Ginolfi}, M., {Jones}, G.~C., {B{\'e}thermin}, M., {et~al.} 2020, \aap, 633,
  A90

\bibitem[{{Grazian} {et~al.}(2020){Grazian}, {Giallongo}, {Fiore}, {Boutsia},
  {Civano}, {Cristiani}, {Cupani}, {Dickinson}, {Fontanot}, {Menci}, \&
  {Romano}}]{Grazian2020}
{Grazian}, A., {Giallongo}, E., {Fiore}, F., {et~al.} 2020, \apj, 897, 94

\bibitem[{{Graziani} {et~al.}(2020){Graziani}, {Schneider}, {Ginolfi}, {Hunt},
  {Maio}, {Glatzle}, \& {Ciardi}}]{Granzini2020}
{Graziani}, L., {Schneider}, R., {Ginolfi}, M., {et~al.} 2020, \mnras, 494,
  1071

\bibitem[{{Gruppioni} {et~al.}(2020){Gruppioni}, {Bethermin}, {Loiacono}, {Le
  Fevre}, {Capak}, {Cassata}, {Faisst}, {Schaerer}, {Silverman}, {Yan},
  {Bardelli}, {Boquien}, {Carraro}, {Cimatti}, {Dessauges-Zavadsky}, {Ginolfi},
  {Fujimoto}, {Hathi}, {Jones}, {Khusanova}, {Koekemoer}, {Lagache}, {Lemaux},
  {Oesch}, {Pozzi}, {Riechers}, {Rodighiero}, {Romano}, {Talia}, {Vallini},
  {Vergani}, {Zamorani}, \& {Zucca}}]{Gruppioni2020}
{Gruppioni}, C., {Bethermin}, M., {Loiacono}, F., {et~al.} 2020, arXiv
  e-prints, arXiv:2006.04974

\bibitem[{{Hashimoto} {et~al.}(2019){Hashimoto}, {Inoue}, {Mawatari}, {Tamura},
  {Matsuo}, {Furusawa}, {Harikane}, {Shibuya}, {Knudsen}, {Kohno}, {Ono},
  {Zackrisson}, {Okamoto}, {Kashikawa}, {Oesch}, {Ouchi}, {Ota}, {Shimizu},
  {Taniguchi}, {Umehata}, \& {Watson}}]{Hashimoto19}
{Hashimoto}, T., {Inoue}, A.~K., {Mawatari}, K., {et~al.} 2019, \pasj, 71, 71

\bibitem[{{Hasinger} {et~al.}(2018){Hasinger}, {Capak}, {Salvato}, {Barger},
  {Cowie}, {Faisst}, {Hemmati}, {Kakazu}, {Kartaltepe}, \&
  {Masters}}]{Haisinger2018}
{Hasinger}, G., {Capak}, P., {Salvato}, M., {et~al.} 2018, \apj, 858, 77

\bibitem[{{Heinis} {et~al.}(2014){Heinis}, {Buat}, {B{\'e}thermin}, {Bock},
  {Burgarella}, {Conley}, {Cooray}, {Farrah}, {Ilbert}, \&
  {Magdis}}]{Heinis2014}
{Heinis}, S., {Buat}, V., {B{\'e}thermin}, M., {et~al.} 2014, \mnras, 437, 1268

\bibitem[{{Hirashita} {et~al.}(2005){Hirashita}, {Nozawa}, {Kozasa}, {Ishii},
  \& {Takeuchi}}]{Hirashita05}
{Hirashita}, H., {Nozawa}, T., {Kozasa}, T., {Ishii}, T.~T., \& {Takeuchi},
  T.~T. 2005, \mnras, 357, 1077

\bibitem[{{Howell} {et~al.}(2010){Howell}, {Armus}, {Mazzarella}, {Evans},
  {Surace}, {Sanders}, {Petric}, {Appleton}, {Bothun}, {Bridge}, {Chan},
  {Charmandaris}, {Frayer}, {Haan}, {Inami}, {Kim}, {Lord}, {Madore},
  {Melbourne}, {Schulz}, {U}, {Vavilkin}, {Veilleux}, \& {Xu}}]{Howell2010}
{Howell}, J.~H., {Armus}, L., {Mazzarella}, J.~M., {et~al.} 2010, \apj, 715,
  572

\bibitem[{{Ilbert} {et~al.}(2006){Ilbert}, {Arnouts}, {McCracken},
  {Bolzonella}, {Bertin}, {Le F{\`e}vre}, {Mellier}, {Zamorani}, {Pell{\`o}},
  {Iovino}, {Tresse}, {Le Brun}, {Bottini}, {Garilli}, {Maccagni}, {Picat},
  {Scaramella}, {Scodeggio}, {Vettolani}, {Zanichelli}, {Adami}, {Bardelli},
  {Cappi}, {Charlot}, {Ciliegi}, {Contini}, {Cucciati}, {Foucaud}, {Franzetti},
  {Gavignaud}, {Guzzo}, {Marano}, {Marinoni}, {Mazure}, {Meneux}, {Merighi},
  {Paltani}, {Pollo}, {Pozzetti}, {Radovich}, {Zucca}, {Bondi}, {Bongiorno},
  {Busarello}, {de La Torre}, {Gregorini}, {Lamareille}, {Mathez}, {Merluzzi},
  {Ripepi}, {Rizzo}, \& {Vergani}}]{Ilbert2006}
{Ilbert}, O., {Arnouts}, S., {McCracken}, H.~J., {et~al.} 2006, \aap, 457, 841

\bibitem[{{Jones} {et~al.}(2020){Jones}, {B{\'e}thermin}, {Fudamoto},
  {Ginolfi}, {Capak}, {Cassata}, {Faisst}, {Le F{\`e}vre}, {Schaerer},
  {Silverman}, {Yan}, {Bardelli}, {Boquien}, {Cimatti}, {Dessauges-Zavadsky},
  {Giavalisco}, {Gruppioni}, {Ibar}, {Khusanova}, {Koekemoer}, {Lemaux},
  {Loiacono}, {Maiolino}, {Oesch}, {Pozzi}, {Riechers}, {Rodighiero}, {Talia},
  {Vallini}, {Vergani}, {Zamorani}, \& {Zucca}}]{Jones2019}
{Jones}, G.~C., {B{\'e}thermin}, M., {Fudamoto}, Y., {et~al.} 2020, \mnras,
  491, L18

\bibitem[{{Kennicutt}(1998)}]{Kennicutt1998}
{Kennicutt}, Robert~C., J. 1998, \araa, 36, 189

\bibitem[{{Kennicutt} \& {Evans}(2012)}]{Kennicutt2012}
{Kennicutt}, R.~C. \& {Evans}, N.~J. 2012, \araa, 50, 531

\bibitem[{{Koprowski} {et~al.}(2020){Koprowski}, {Coppin}, {Geach},
  {Dudzevi{\v{c}}i{\={u}}t{\.{e}}}, {Smail}, {Almaini}, {An}, {Blain},
  {Chapman}, {Chen}, {Conselice}, {Dunlop}, {Farrah}, {Gullberg}, {Hartley},
  {Ivison}, {Karska}, {Maltby}, {Malek}, {Micha{\l}owski}, {Pope}, {Salim},
  {Scott}, {Simpson}, {Simpson}, {Swinbank}, {Thomson}, {Wardlow}, {van der
  Werf}, \& {Whitaker}}]{Koprowski2020}
{Koprowski}, M.~P., {Coppin}, K.~E.~K., {Geach}, J.~E., {et~al.} 2020, \mnras,
  492, 4927

\bibitem[{{Koprowski} {et~al.}(2018){Koprowski}, {Coppin}, {Geach}, {McLure},
  {Almaini}, {Blain}, {Bremer}, {Bourne}, {Chapman}, {Conselice}, {Dunlop},
  {Farrah}, {Hartley}, {Karim}, {Knudsen}, {Micha{\l}owski}, {Scott},
  {Simpson}, {Smith}, \& {van der Werf}}]{Koprowski2018}
{Koprowski}, M.~P., {Coppin}, K.~E.~K., {Geach}, J.~E., {et~al.} 2018, \mnras
  [\eprint[arXiv]{1801.00791}]

\bibitem[{{Laporte} {et~al.}(2017){Laporte}, {Ellis}, {Boone}, {Bauer},
  {Qu{\'e}nard}, {Roberts-Borsani}, {Pell{\'o}}, {P{\'e}rez-Fournon}, \&
  {Streblyanska}}]{Laporte17}
{Laporte}, N., {Ellis}, R.~S., {Boone}, F., {et~al.} 2017, \apjl, 837, L21

\bibitem[{{Le F{\`e}vre} {et~al.}(2019){Le F{\`e}vre}, {B{\'e}thermin},
  {Faisst}, {Capak}, {Cassata}, {Silverman}, {Schaerer}, \&
  {Yan}}]{Lefevre2019}
{Le F{\`e}vre}, O., {B{\'e}thermin}, M., {Faisst}, A., {et~al.} 2019, arXiv
  e-prints, arXiv:1910.09517

\bibitem[{{Le F{\`e}vre} {et~al.}(2015){Le F{\`e}vre}, {Tasca}, {Cassata},
  {Garilli}, {Le Brun}, {Maccagni}, {Pentericci}, {Thomas}, {Vanzella}, \&
  {Zamorani}}]{Lefevre2015}
{Le F{\`e}vre}, O., {Tasca}, L.~A.~M., {Cassata}, P., {et~al.} 2015, \aap, 576,
  A79

\bibitem[{{Lee} {et~al.}(2012){Lee}, {Ferguson}, {Wiklind}, {Dahlen},
  {Dickinson}, {Giavalisco}, {Grogin}, {Papovich}, {Messias}, \&
  {Guo}}]{Lee2012}
{Lee}, K.-S., {Ferguson}, H.~C., {Wiklind}, T., {et~al.} 2012, \apj, 752, 66

\bibitem[{{Liu} {et~al.}(2019){Liu}, {Lang}, {Magnelli}, {Schinnerer},
  {Leslie}, {Fudamoto}, {Bondi}, {Groves}, {Jim{\'e}nez-Andrade}, {Harrington},
  {Karim}, {Oesch}, {Sargent}, {Vardoulaki}, {B{\v{a}}descu}, {Moser},
  {Bertoldi}, {Battisti}, {da Cunha}, {Zavala}, {Vaccari}, {Davidzon},
  {Riechers}, \& {Aravena}}]{Liu2019}
{Liu}, D., {Lang}, P., {Magnelli}, B., {et~al.} 2019, \apjs, 244, 40

\bibitem[{{Lutz} {et~al.}(2011){Lutz}, {Poglitsch}, {Altieri}, {Andreani},
  {Aussel}, {Berta}, {Bongiovanni}, {Brisbin}, {Cava}, {Cepa}, {Cimatti},
  {Daddi}, {Dominguez-Sanchez}, {Elbaz}, {F{\"o}rster Schreiber}, {Genzel},
  {Grazian}, {Gruppioni}, {Harwit}, {Le Floc'h}, {Magdis}, {Magnelli},
  {Maiolino}, {Nordon}, {P{\'e}rez Garc{\'\i}a}, {Popesso}, {Pozzi},
  {Riguccini}, {Rodighiero}, {Saintonge}, {Sanchez Portal}, {Santini}, {Shao},
  {Sturm}, {Tacconi}, {Valtchanov}, {Wetzstein}, \& {Wieprecht}}]{Lutz2011}
{Lutz}, D., {Poglitsch}, A., {Altieri}, B., {et~al.} 2011, \aap, 532, A90

\bibitem[{{Madau} \& {Dickinson}(2014)}]{Madau2014}
{Madau}, P. \& {Dickinson}, M. 2014, \araa, 52, 415

\bibitem[{{Magdis} {et~al.}(2012){Magdis}, {Daddi}, {B{\'e}thermin}, {Sargent},
  {Elbaz}, {Pannella}, {Dickinson}, {Dannerbauer}, {da Cunha}, {Walter},
  {Rigopoulou}, {Charmandaris}, {Hwang}, \& {Kartaltepe}}]{Magdis2012}
{Magdis}, G.~E., {Daddi}, E., {B{\'e}thermin}, M., {et~al.} 2012, \apj, 760, 6

\bibitem[{{Maiolino} {et~al.}(2004){Maiolino}, {Schneider}, {Oliva}, {Bianchi},
  {Ferrara}, {Mannucci}, {Pedani}, \& {Roca Sogorb}}]{Maiolino2004}
{Maiolino}, R., {Schneider}, R., {Oliva}, E., {et~al.} 2004, \nat, 431, 533

\bibitem[{{McLure} {et~al.}(2018){McLure}, {Dunlop}, {Cullen}, {Bourne},
  {Best}, {Khochfar}, {Bowler}, {Biggs}, {Geach}, {Scott}, {Micha{\l}owski},
  {Rujopakarn}, {van Kampen}, {Kirkpatrick}, \& {Pope}}]{Mclure2018}
{McLure}, R.~J., {Dunlop}, J.~S., {Cullen}, F., {et~al.} 2018, \mnras, 476,
  3991

\bibitem[{{McMullin} {et~al.}(2007){McMullin}, {Waters}, {Schiebel}, {Young},
  \& {Golap}}]{Mcmullin2007}
{McMullin}, J.~P., {Waters}, B., {Schiebel}, D., {Young}, W., \& {Golap}, K.
  2007, in Astronomical Society of the Pacific Conference Series, Vol. 376,
  Astronomical Data Analysis Software and Systems XVI, ed. R.~A. {Shaw},
  F.~{Hill}, \& D.~J. {Bell}, 127

\bibitem[{{Meurer} {et~al.}(1999){Meurer}, {Heckman}, \&
  {Calzetti}}]{Meurer1999}
{Meurer}, G.~R., {Heckman}, T.~M., \& {Calzetti}, D. 1999, \apj, 521, 64

\bibitem[{{Narayanan} {et~al.}(2018){Narayanan}, {Dav{\'e}}, {Johnson},
  {Thompson}, {Conroy}, \& {Geach}}]{Narayanan2018}
{Narayanan}, D., {Dav{\'e}}, R., {Johnson}, B.~D., {et~al.} 2018, \mnras, 474,
  1718

\bibitem[{{Nozawa} {et~al.}(2003){Nozawa}, {Kozasa}, {Umeda}, {Maeda}, \&
  {Nomoto}}]{Nozawa03}
{Nozawa}, T., {Kozasa}, T., {Umeda}, H., {Maeda}, K., \& {Nomoto}, K. 2003,
  \apj, 598, 785

\bibitem[{{Oesch} {et~al.}(2018){Oesch}, {Bouwens}, {Illingworth}, {Labb{\'e}},
  \& {Stefanon}}]{Oesch2018}
{Oesch}, P.~A., {Bouwens}, R.~J., {Illingworth}, G.~D., {Labb{\'e}}, I., \&
  {Stefanon}, M. 2018, \apj, 855, 105

\bibitem[{{Oliver} {et~al.}(2012){Oliver}, {Bock}, {Altieri}, {Amblard},
  {Arumugam}, {Aussel}, {Babbedge}, {Beelen}, {B{\'e}thermin}, {Blain},
  {Boselli}, {Bridge}, {Brisbin}, {Buat}, {Burgarella},
  {Castro-Rodr{\'\i}guez}, {Cava}, {Chanial}, {Cirasuolo}, {Clements},
  {Conley}, {Conversi}, {Cooray}, {Dowell}, {Dubois}, {Dwek}, {Dye}, {Eales},
  {Elbaz}, {Farrah}, {Feltre}, {Ferrero}, {Fiolet}, {Fox}, {Franceschini},
  {Gear}, {Giovannoli}, {Glenn}, {Gong}, {Gonz{\'a}lez Solares}, {Griffin},
  {Halpern}, {Harwit}, {Hatziminaoglou}, {Heinis}, {Hurley}, {Hwang}, {Hyde},
  {Ibar}, {Ilbert}, {Isaak}, {Ivison}, {Lagache}, {Le Floc'h}, {Levenson},
  {Faro}, {Lu}, {Madden}, {Maffei}, {Magdis}, {Mainetti}, {Marchetti},
  {Marsden}, {Marshall}, {Mortier}, {Nguyen}, {O'Halloran}, {Omont}, {Page},
  {Panuzzo}, {Papageorgiou}, {Patel}, {Pearson}, {P{\'e}rez-Fournon}, {Pohlen},
  {Rawlings}, {Raymond}, {Rigopoulou}, {Riguccini}, {Rizzo}, {Rodighiero},
  {Roseboom}, {Rowan-Robinson}, {S{\'a}nchez Portal}, {Schulz}, {Scott},
  {Seymour}, {Shupe}, {Smith}, {Stevens}, {Symeonidis}, {Trichas}, {Tugwell},
  {Vaccari}, {Valtchanov}, {Vieira}, {Viero}, {Vigroux}, {Wang}, {Ward},
  {Wardlow}, {Wright}, {Xu}, \& {Zemcov}}]{Oliver2012}
{Oliver}, S.~J., {Bock}, J., {Altieri}, B., {et~al.} 2012, \mnras, 424, 1614

\bibitem[{{Overzier} {et~al.}(2011){Overzier}, {Heckman}, {Wang}, {Armus},
  {Buat}, {Howell}, {Meurer}, {Seibert}, {Siana}, {Basu-Zych}, {Charlot},
  {Gon{\c c}alves}, {Martin}, {Neill}, {Rich}, {Salim}, \&
  {Schiminovich}}]{Overzier2011}
{Overzier}, R.~A., {Heckman}, T.~M., {Wang}, J., {et~al.} 2011, \apjl, 726, L7

\bibitem[{{Pannella} {et~al.}(2015){Pannella}, {Elbaz}, {Daddi}, {Dickinson},
  {Hwang}, {Schreiber}, {Strazzullo}, {Aussel}, {Bethermin}, {Buat},
  {Charmandaris}, {Cibinel}, {Juneau}, {Ivison}, {Le Borgne}, {Le Floc'h},
  {Leiton}, {Lin}, {Magdis}, {Morrison}, {Mullaney}, {Onodera}, {Renzini},
  {Salim}, {Sargent}, {Scott}, {Shu}, \& {Wang}}]{Pannella2015}
{Pannella}, M., {Elbaz}, D., {Daddi}, E., {et~al.} 2015, \apj, 807, 141

\bibitem[{{Popping} {et~al.}(2017){Popping}, {Puglisi}, \&
  {Norman}}]{Popping2017}
{Popping}, G., {Puglisi}, A., \& {Norman}, C.~A. 2017, \mnras, 472, 2315

\bibitem[{{Prevot} {et~al.}(1984){Prevot}, {Lequeux}, {Maurice}, {Prevot}, \&
  {Rocca-Volmerange}}]{Prevot1984}
{Prevot}, M.~L., {Lequeux}, J., {Maurice}, E., {Prevot}, L., \&
  {Rocca-Volmerange}, B. 1984, \aap, 132, 389

\bibitem[{{Reddy} {et~al.}(2018){Reddy}, {Oesch}, {Bouwens}, {Montes},
  {Illingworth}, {Steidel}, {van Dokkum}, {Atek}, {Carollo}, {Cibinel},
  {Holden}, {Labb{\'e}}, {Magee}, {Morselli}, {Nelson}, \&
  {Wilkins}}]{Reddy2018}
{Reddy}, N.~A., {Oesch}, P.~A., {Bouwens}, R.~J., {et~al.} 2018, \apj, 853, 56

\bibitem[{{Salim} \& {Narayanan}(2020)}]{Salim2020}
{Salim}, S. \& {Narayanan}, D. 2020, arXiv e-prints, arXiv:2001.03181

\bibitem[{{Santini} {et~al.}(2014){Santini}, {Maiolino}, {Magnelli}, {Lutz},
  {Lamastra}, {Li Causi}, {Eales}, {Andreani}, {Berta}, {Buat}, {Cooray},
  {Cresci}, {Daddi}, {Farrah}, {Fontana}, {Franceschini}, {Genzel}, {Granato},
  {Grazian}, {Le Floc'h}, {Magdis}, {Magliocchetti}, {Mannucci}, {Menci},
  {Nordon}, {Oliver}, {Popesso}, {Pozzi}, {Riguccini}, {Rodighiero}, {Rosario},
  {Salvato}, {Scott}, {Silva}, {Tacconi}, {Viero}, {Wang}, {Wuyts}, \&
  {Xu}}]{Santiani2014}
{Santini}, P., {Maiolino}, R., {Magnelli}, B., {et~al.} 2014, \aap, 562, A30

\bibitem[{{Schaerer} {et~al.}(2020){Schaerer}, {Ginolfi}, {Bethermin},
  {Fudamoto}, {Oesch}, {Le Fevre}, {Faisst}, {Capak}, {Cassata}, {Silverman},
  {Yan}, {Jones}, {Amorin}, {Bardelli}, {Boquien}, {Cimatti},
  {Dessauges-Zavadsky}, {Giavalisco}, {Hathi}, {Fujimoto}, {Ibar}, {Koekemoer},
  {Lagache}, {Lemaux}, {Loiacono}, {Maiolino}, {Narayanan}, {Morselli},
  {Mendez-Hernandez}, {Pozzi}, {Riechers}, {Talia}, {Toft}, {Vallini},
  {Vergani}, {Zamorani}, \& {Zucca}}]{Schaerer2020}
{Schaerer}, D., {Ginolfi}, M., {Bethermin}, M., {et~al.} 2020, arXiv e-prints,
  arXiv:2002.00979

\bibitem[{{Schneider} {et~al.}(2004){Schneider}, {Ferrara}, \&
  {Salvaterra}}]{Schneider04}
{Schneider}, R., {Ferrara}, A., \& {Salvaterra}, R. 2004, \mnras, 351, 1379

\bibitem[{{Schreiber} {et~al.}(2018){Schreiber}, {Elbaz}, {Pannella}, {Ciesla},
  {Wang}, \& {Franco}}]{Schreiber2018}
{Schreiber}, C., {Elbaz}, D., {Pannella}, M., {et~al.} 2018, \aap, 609, A30

\bibitem[{{Schreiber} {et~al.}(2015){Schreiber}, {Pannella}, {Elbaz},
  {B{\'e}thermin}, {Inami}, {Dickinson}, {Magnelli}, {Wang}, {Aussel}, {Daddi},
  {Juneau}, {Shu}, {Sargent}, {Buat}, {Faber}, {Ferguson}, {Giavalisco},
  {Koekemoer}, {Magdis}, {Morrison}, {Papovich}, {Santini}, \&
  {Scott}}]{Schreiber2015}
{Schreiber}, C., {Pannella}, M., {Elbaz}, D., {et~al.} 2015, \aap, 575, A74

\bibitem[{{Smit} {et~al.}(2018){Smit}, {Bouwens}, {Carniani}, {Oesch},
  {Labb{\'e}}, {Illingworth}, {van der Werf}, {Bradley}, {Gonzalez}, {Hodge},
  {Holwerda}, {Maiolino}, \& {Zheng}}]{Smit18}
{Smit}, R., {Bouwens}, R.~J., {Carniani}, S., {et~al.} 2018, \nat, 553, 178

\bibitem[{{Stanway} {et~al.}(2016){Stanway}, {Eldridge}, \&
  {Becker}}]{Stanway2016}
{Stanway}, E.~R., {Eldridge}, J.~J., \& {Becker}, G.~D. 2016, \mnras, 456, 485

\bibitem[{{Steinhardt} {et~al.}(2014){Steinhardt}, {Speagle}, {Capak},
  {Silverman}, {Carollo}, {Dunlop}, {Hashimoto}, {Hsieh}, {Ilbert}, \& {Le
  Fevre}}]{Steinhardt2014}
{Steinhardt}, C.~L., {Speagle}, J.~S., {Capak}, P., {et~al.} 2014, \apjl, 791,
  L25

\bibitem[{{Stratta} {et~al.}(2007){Stratta}, {Maiolino}, {Fiore}, \&
  {D'Elia}}]{Stratta2007}
{Stratta}, G., {Maiolino}, R., {Fiore}, F., \& {D'Elia}, V. 2007, \apjl, 661,
  L9

\bibitem[{{Takeuchi} {et~al.}(2012){Takeuchi}, {Yuan}, {Ikeyama}, {Murata}, \&
  {Inoue}}]{Takeuchi2012}
{Takeuchi}, T.~T., {Yuan}, F.-T., {Ikeyama}, A., {Murata}, K.~L., \& {Inoue},
  A.~K. 2012, \apj, 755, 144

\bibitem[{{Tamura} {et~al.}(2019){Tamura}, {Mawatari}, {Hashimoto}, {Inoue},
  {Zackrisson}, {Christensen}, {Binggeli}, {Matsuda}, {Matsuo}, {Takeuchi},
  {Asano}, {Sunaga}, {Shimizu}, {Okamoto}, {Yoshida}, {Lee}, {Shibuya},
  {Taniguchi}, {Umehata}, {Hatsukade}, {Kohno}, \& {Ota}}]{Tamura19}
{Tamura}, Y., {Mawatari}, K., {Hashimoto}, T., {et~al.} 2019, \apj, 874, 27

\bibitem[{{Todini} \& {Ferrara}(2001)}]{Todini01}
{Todini}, P. \& {Ferrara}, A. 2001, \mnras, 325, 726

\bibitem[{{Wang} {et~al.}(2019){Wang}, {Schreiber}, {Elbaz}, {Yoshimura},
  {Kohno}, {Shu}, {Yamaguchi}, {Pannella}, {Franco}, {Huang}, {Lim}, \&
  {Wang}}]{Wang19}
{Wang}, T., {Schreiber}, C., {Elbaz}, D., {et~al.} 2019, \nat, 572, 211

\bibitem[{{Watson} {et~al.}(2015){Watson}, {Christensen}, {Knudsen}, {Richard},
  {Gallazzi}, \& {Micha{\l}owski}}]{Watson15}
{Watson}, D., {Christensen}, L., {Knudsen}, K.~K., {et~al.} 2015, \nat, 519,
  327

\bibitem[{{Whitaker} {et~al.}(2017){Whitaker}, {Pope}, {Cybulski}, {Casey},
  {Popping}, \& {Yun}}]{Whitaker2017}
{Whitaker}, K.~E., {Pope}, A., {Cybulski}, R., {et~al.} 2017, The Astrophysical
  Journal, 850, 208

\bibitem[{{Wilkins} {et~al.}(2011){Wilkins}, {Bunker}, {Stanway}, {Lorenzoni},
  \& {Caruana}}]{Wilkins2011}
{Wilkins}, S.~M., {Bunker}, A.~J., {Stanway}, E., {Lorenzoni}, S., \&
  {Caruana}, J. 2011, \mnras, 417, 717

\bibitem[{{Wilkins} {et~al.}(2018){Wilkins}, {Feng}, {Di Matteo}, {Croft},
  {Lovell}, \& {Thomas}}]{Wilkins2018}
{Wilkins}, S.~M., {Feng}, Y., {Di Matteo}, T., {et~al.} 2018, \mnras, 473, 5363

\bibitem[{{Wilkins} {et~al.}(2019){Wilkins}, {Lovell}, \&
  {Stanway}}]{Wilkins19}
{Wilkins}, S.~M., {Lovell}, C.~C., \& {Stanway}, E.~R. 2019, \mnras, 490, 5359

\bibitem[{{Wilkins} {et~al.}(2008){Wilkins}, {Trentham}, \&
  {Hopkins}}]{Wilkins08}
{Wilkins}, S.~M., {Trentham}, N., \& {Hopkins}, A.~M. 2008, \mnras, 385, 687

\bibitem[{{Williams} {et~al.}(2019){Williams}, {Labbe}, {Spilker}, {Stefanon},
  {Leja}, {Whitaker}, {Bezanson}, {Narayanan}, {Oesch}, \&
  {Weiner}}]{Williams19}
{Williams}, C.~C., {Labbe}, I., {Spilker}, J., {et~al.} 2019, \apj, 884, 154

\end{thebibliography}
%

\begin{appendix}
\onecolumn

\section{Measured and Predicted Infrared Luminosities}
\label{appendix:LIR}


Using the best fit IRX-$\beta$ relations (Equation \ref{eqn:irxb} and Table \ref{table:IRXb}), we estimated the IR luminosity of the galaxies observed in the ALPINE programme (Table \ref{table:A_LIR}). While the IRX-$\beta$ assuming $\beta_0=-2.62$ is our fiducial relation (Section \ref{sec:IRX-beta}), we list $L_{\rm{IR}}$ estimated using the IRX-$\beta$ relation with $\beta_0=-2.32$.
When a $\beta$ is bluer than $\beta_0$ for each case, the galaxy is consistent to be dust-free, meaning that the estimated $L_{\mathrm{IR}}$ is consistent with zero.
The $L_{\rm{IR}}$ estimated with our fiducial IRX-$\beta$ relation are used to estimate the total star formation rates in \citet{Schaerer2020}.

\renewcommand{\arraystretch}{0.91} 
\begin{longtable}{lcccccc}
\caption{\label{table:A_LIR} Summary of the total IR luminosity ($\lambda=8-1000\,\rm{\mu m}$) estimations of our sample. $L_{\rm{IR,Meas}}$ is the measured IR luminosity for detections and $3\,\sigma$ upperlimits for nondetections. The $L_{\rm{IR,Meas}}$ is calculated using the FIR SED template from \citet{Bethermin2019}. The $L_{\rm{IR, -2.23}}$ and the $L_{\rm{IR, -2.62}}$ is the estimated total infrared luminosity using our best fit IRX-$\beta$ relations assuming $\beta_0=-2.23$ and $\beta_0=-2.62$ (our fiducial value), respectively.
The ``--'' sign indicates that the UV slope of the galaxy is bluer than the assumed intrinsic $\beta$ (i.e., the estimated $L_{\mathrm{IR}}$ is consistent with zero).}\\
\hline\hline
Name & z & $\beta$ & $L_{\rm{IR,Meas}}$ & $L_{\rm{IR, -2.23}}$ & $L_{\rm{IR, -2.62}}$ \\ 
& & &  $\rm{log(L/L_{\odot})}$ & $\rm{log(L/L_{\odot})}$ & $\rm{log(L/L_{\odot})}$ \\
\hline
\endfirsthead
\caption{continued.}\\
\hline\hline
Name & z & $\beta$ & $L_{\rm UV}$ & $L_{\rm{IR,Meas}}$ & $L_{\rm{IR, -2.23}}$ & $L_{\rm{IR, -2.62}}$\\ 
& & &  $\rm{log(L/L_{\odot})}$ & $\rm{log(L/L_{\odot})}$ & $\rm{log(L/L_{\odot})}$ \\
\hline
\endhead
\hline
\endfoot
CANDELS\_GOODSS\_12 & $4.43$ & $ -2.04$ & $<11.49$ & $10.36$ & $10.74$\\
CANDELS\_GOODSS\_19 & $4.50$ & $-0.60$ & $11.57$ & $11.45$ & $11.34$\\
CANDELS\_GOODSS\_21 & $5.58$ & $ -1.43$ & $<11.37$ & $10.61$ & $10.59$\\
CANDELS\_GOODSS\_32 & $4.41$ & $-0.86$ & $11.47$ & $11.17$ & $11.09$\\
CANDELS\_GOODSS\_37 & $4.52$ & $ -1.40$ & $<11.48$ & $10.06$ & $10.08$\\
CANDELS\_GOODSS\_38 & $5.57$ & $ -1.58$ & $<11.37$ & $10.84$ & $10.86$\\
CANDELS\_GOODSS\_42 & $5.54$ & $ -2.04$ & $<11.28$ & $10.07$ & $10.39$\\
CANDELS\_GOODSS\_47 & $5.58$ & $ -1.81$ & $<11.46$ & $10.42$ & $10.53$\\
CANDELS\_GOODSS\_57 & $5.56$ & $ -1.19$ & $<11.39$ & $11.06$ & $11.00$\\
CANDELS\_GOODSS\_75 & $5.60$ & $ -2.31$ & $<11.29$ & -- & $9.86$\\
CANDELS\_GOODSS\_8 & $5.52$ & $ -1.58$ & $<11.33$ & $10.74$ & $10.76$\\
DEIMOS\_COSMOS\_206253 & $4.47$ & $ -1.51$ & $<11.46$ & $10.97$ & $11.02$\\
DEIMOS\_COSMOS\_224751 & $5.72$ & $ -2.66$ & $<11.12$ & -- & --\\
DEIMOS\_COSMOS\_274035 & $4.48$ & $ -2.16$ & $<11.59$ & $10.05$ & $10.74$\\
DEIMOS\_COSMOS\_298678 & $5.68$ & $ -1.83$ & $<11.44$ & $10.45$ & $10.58$\\
DEIMOS\_COSMOS\_308643 & $4.53$ & $ -1.84$ & $<11.62$ & $11.11$ & $11.29$\\
DEIMOS\_COSMOS\_328419 & $5.72$ & $ -1.84$ & $<11.54$ & $10.55$ & $10.68$\\
DEIMOS\_COSMOS\_336830 & $5.71$ & $ -2.59$ & $<11.33$ & -- & $9.03$\\
DEIMOS\_COSMOS\_351640 & $5.71$ & $ -1.97$ & $<11.39$ & $10.68$ & $10.91$\\
DEIMOS\_COSMOS\_357722 & $5.74$ & $ -2.24$ & $<11.28$ & -- & $10.35$\\
DEIMOS\_COSMOS\_372292 & $5.14$ & $ -1.72$ & $<11.37$ & $10.82$ & $10.89$\\
DEIMOS\_COSMOS\_378903 & $5.43$ & $ -2.36$ & $<11.45$ & -- & $9.96$\\
DEIMOS\_COSMOS\_396844 & $4.54$ & $-1.38$ & $11.67$ & $11.21$ & $11.23$\\
DEIMOS\_COSMOS\_400160 & $4.53$ & $ -1.87$ & $<11.70$ & $11.03$ & $11.23$\\
DEIMOS\_COSMOS\_403030 & $4.57$ & $ -1.77$ & $<11.67$ & $10.91$ & $11.05$\\
DEIMOS\_COSMOS\_406956 & $5.68$ & $ -2.08$ & $<11.34$ & $10.11$ & $10.51$\\
DEIMOS\_COSMOS\_412589 & $4.43$ & $ -2.66$ & $<11.56$ & -- & -- \\
DEIMOS\_COSMOS\_416105 & $5.63$ & $ -2.28$ & $<11.18$ & -- & $10.44$\\
DEIMOS\_COSMOS\_417567 & $5.67$ & $-1.87$ & $11.58$ & $11.14$ & $11.28$\\
DEIMOS\_COSMOS\_420065 & $5.73$ & $ -2.17$ & $<11.32$ & $9.65$ & $10.36$\\
DEIMOS\_COSMOS\_421062 & $5.58$ & $ -1.51$ & $<11.16$ & $11.15$ & $11.15$\\
DEIMOS\_COSMOS\_422677 & $4.44$ & $-1.24$ & $11.69$ & $11.38$ & $11.36$\\
DEIMOS\_COSMOS\_430951 & $5.68$ & $ -2.39$ & $<11.67$ & -- & $10.33$\\
DEIMOS\_COSMOS\_431067 & $4.43$ & $ -1.73$ & $<11.44$ & $10.61$ & $10.74$\\
DEIMOS\_COSMOS\_432340 & $4.41$ & $ -1.63$ & $<11.66$ & $11.29$ & $11.38$\\
DEIMOS\_COSMOS\_434239 & $4.49$ & $ -1.31$ & $<11.48$ & $11.40$ & $11.40$\\
DEIMOS\_COSMOS\_442844 & $4.49$ & $ -2.40$ & $<11.47$ & -- & $10.11$\\
DEIMOS\_COSMOS\_454608 & $4.58$ & $ -1.63$ & $<11.57$ & $11.08$ & $11.16$\\
DEIMOS\_COSMOS\_460378 & $5.39$ & $-1.44$ & $11.31$ & $11.13$ & $11.11$\\
DEIMOS\_COSMOS\_470116 & $5.68$ & $ -2.36$ & $<11.35$ & -- & $10.01$\\
DEIMOS\_COSMOS\_471063 & $5.72$ & $ -2.23$ & $<11.58$ & $7.80$ & $10.04$\\
DEIMOS\_COSMOS\_472215 & $5.64$ & $ -0.79$ & $<11.53$ & $8.48$ & $8.36$\\
DEIMOS\_COSMOS\_488399 & $5.68$ & $-1.88$ & $11.67$ & $10.81$ & $10.97$\\
DEIMOS\_COSMOS\_493583 & $4.52$ & $-2.01$ & $11.50$ & $10.56$ & $10.90$\\
DEIMOS\_COSMOS\_494057 & $5.54$ & $-1.88$ & $11.51$ & $11.00$ & $11.15$\\
DEIMOS\_COSMOS\_494763 & $5.24$ & $ -1.48$ & $<11.40$ & $10.80$ & $10.80$\\
DEIMOS\_COSMOS\_503575 & $5.65$ & $ -1.75$ & $<11.31$ & $9.88$ & $9.97$\\
DEIMOS\_COSMOS\_510660 & $4.55$ & $ -2.11$ & $<11.52$ & $10.21$ & $10.73$\\
DEIMOS\_COSMOS\_519281 & $5.57$ & $ -1.88$ & $<11.51$ & $10.71$ & $10.87$\\
DEIMOS\_COSMOS\_536534 & $5.69$ & $ -1.98$ & $<11.42$ & $10.81$ & $11.06$\\
DEIMOS\_COSMOS\_539609 & $5.17$ & $-2.21$ & $11.48$ & $9.70$ & $10.79$\\
DEIMOS\_COSMOS\_549131 & $5.55$ & $ -2.22$ & $<11.44$ & $9.19$ & $10.63$\\
DEIMOS\_COSMOS\_550156 & $4.42$ & $ -1.87$ & $<11.46$ & $10.74$ & $10.94$\\
DEIMOS\_COSMOS\_552206 & $5.51$ & $-0.98$ & $11.71$ & $11.63$ & $11.52$\\
DEIMOS\_COSMOS\_567070 & $4.56$ & $ -1.98$ & $<11.58$ & $10.57$ & $10.87$\\
DEIMOS\_COSMOS\_576372 & $5.66$ & $ -2.66$ & $<11.45$ & -- & --\\
DEIMOS\_COSMOS\_586681 & $5.87$ & $ -1.87$ & $<11.55$ & $10.75$ & $10.90$\\
DEIMOS\_COSMOS\_592644 & $4.53$ & $ -2.16$ & $<11.46$ & $10.07$ & $10.76$\\
DEIMOS\_COSMOS\_627939 & $4.53$ & $ -1.27$ & $<11.43$ & $11.33$ & $11.32$\\
DEIMOS\_COSMOS\_628063 & $4.54$ & $ -1.60$ & $<11.56$ & $10.89$ & $10.97$\\
DEIMOS\_COSMOS\_628137 & $5.68$ & $ -2.57$ & $<11.31$ & -- & $9.01$\\
DEIMOS\_COSMOS\_629750 & $5.12$ & $ -1.72$ & $<11.45$ & $10.60$ & $10.67$\\
DEIMOS\_COSMOS\_630594 & $4.45$ & $ -1.44$ & $<11.50$ & $11.09$ & $11.13$\\
DEIMOS\_COSMOS\_665509 & $4.53$ & $ -1.93$ & $<11.45$ & $10.81$ & $11.06$\\
DEIMOS\_COSMOS\_665626 & $4.58$ & $ -2.18$ & $<11.46$ & $9.24$ & $10.07$\\
DEIMOS\_COSMOS\_680104 & $4.53$ & $ -2.23$ & $<11.59$ & $8.32$ & $10.61$\\
DEIMOS\_COSMOS\_683613 & $5.54$ & $-1.30$ & $11.65$ & $11.16$ & $11.12$\\
DEIMOS\_COSMOS\_709575 & $4.42$ & $ -1.50$ & $<11.63$ & $11.04$ & $11.09$\\
DEIMOS\_COSMOS\_722679 & $5.76$ & $ -2.66$ & $<11.38$ & -- & --\\
DEIMOS\_COSMOS\_733857 & $4.55$ & $ -1.75$ & $<11.56$ & $11.08$ & $11.22$\\
DEIMOS\_COSMOS\_742174 & $5.64$ & $ -2.13$ & $<11.40$ & $10.04$ & $10.56$\\
DEIMOS\_COSMOS\_743730 & $4.52$ & $ -1.02$ & $<11.48$ & $10.85$ & $10.80$\\
DEIMOS\_COSMOS\_761315 & $4.58$ & $ -1.86$ & $<11.50$ & $10.48$ & $10.68$\\
DEIMOS\_COSMOS\_773957 & $5.68$ & $ -2.05$ & $<11.35$ & $10.38$ & $10.71$\\
DEIMOS\_COSMOS\_787780 & $4.51$ & $ -1.32$ & $<11.49$ & $10.89$ & $10.89$\\
DEIMOS\_COSMOS\_790930 & $5.69$ & $ -2.66$ & $<11.26$ & -- & --\\
DEIMOS\_COSMOS\_803480 & $4.54$ & $ -2.44$ & $<11.42$ & -- & $10.21$\\
DEIMOS\_COSMOS\_814483 & $4.58$ & $ -1.87$ & $<11.62$ & $11.04$ & $11.24$\\
DEIMOS\_COSMOS\_818760 & $4.55$ & $-0.55$ & $12.16$ & $11.96$ & $11.84$\\
DEIMOS\_COSMOS\_834764 & $4.50$ & $ -1.99$ & $<11.65$ & $10.75$ & $11.07$\\
DEIMOS\_COSMOS\_838532 & $4.53$ & $ -2.22$ & $<11.62$ & $9.08$ & $10.57$\\
DEIMOS\_COSMOS\_842313 & $4.55$ & $ -1.44$ & $<13.05$ & $11.81$ & $11.84$\\
DEIMOS\_COSMOS\_843045 & $5.82$ & $ -1.84$ & $<11.59$ & $10.75$ & $10.88$\\
DEIMOS\_COSMOS\_848185 & $5.28$ & $-1.14$ & $11.73$ & $11.59$ & $11.51$\\
DEIMOS\_COSMOS\_859732 & $4.53$ & $ -1.74$ & $<11.48$ & $10.68$ & $10.81$\\
DEIMOS\_COSMOS\_869970 & $5.20$ & $ -2.46$ & $<11.33$ & -- & $9.97$\\
DEIMOS\_COSMOS\_873321 & $5.16$ & $ -1.48$ & $<11.54$ & $11.31$ & $11.30$\\
DEIMOS\_COSMOS\_873756 & $4.55$ & $-1.59$ & $12.26$ & $10.88$ & $10.96$\\
DEIMOS\_COSMOS\_880016 & $4.54$ & $ -1.38$ & $<11.52$ & $11.02$ & $11.04$\\
DEIMOS\_COSMOS\_881725 & $4.58$ & $-1.20$ & $12.23$ & $11.36$ & $11.34$\\
DEIMOS\_COSMOS\_910650 & $5.66$ & $ -2.13$ & $<11.40$ & $9.91$ & $10.45$\\
DEIMOS\_COSMOS\_920848 & $4.55$ & $ -1.72$ & $<11.53$ & $10.75$ & $10.87$\\
DEIMOS\_COSMOS\_926434 & $4.45$ & $ -1.67$ & $<11.56$ & $11.34$ & $11.44$\\
DEIMOS\_COSMOS\_933876 & $4.42$ & $ -1.97$ & $<11.62$ & $10.50$ & $10.78$\\
vuds\_cosmos\_5100537582 & $4.55$ & $ -2.06$ & $<11.55$ & $10.25$ & $10.66$\\
vuds\_cosmos\_5100541407 & $4.55$ & $ -1.26$ & $<11.70$ & $11.18$ & $11.17$\\
vuds\_cosmos\_5100559223 & $4.56$ & $ -1.42$ & $<11.66$ & $11.11$ & $11.13$\\
vuds\_cosmos\_5100822662 & $4.52$ & $-1.32$ & $11.45$ & $11.39$ & $11.39$\\
vuds\_cosmos\_5100969402 & $4.59$ & $-1.94$ & $11.65$ & $10.66$ & $10.92$\\
vuds\_cosmos\_5100994794 & $4.58$ & $-1.63$ & $11.20$ & $10.94$ & $11.03$\\
vuds\_cosmos\_5101013812 & $4.42$ & $ -1.91$ & $<11.41$ & $10.71$ & $10.94$\\
vuds\_cosmos\_5101209780 & $4.57$ & $-1.92$ & $12.04$ & $10.95$ & $11.18$\\
vuds\_cosmos\_5101210235 & $4.57$ & $ -2.06$ & $<11.40$ & $10.68$ & $11.08$\\
vuds\_cosmos\_5101218326 & $4.57$ & $-0.86$ & $11.79$ & $11.92$ & $11.84$\\
vuds\_cosmos\_5101244930 & $4.58$ & $ -1.87$ & $<11.68$ & $10.86$ & $11.06$\\
vuds\_cosmos\_5101288969 & $5.70$ & $ -2.08$ & $<11.23$ & $10.38$ & $10.78$\\
vuds\_cosmos\_510148750 & $4.51$ & $ -2.41$ & $<11.33$ & -- & $10.18$\\
vuds\_cosmos\_510327576 & $4.56$ & $ -2.35$ & $<11.64$ & -- & $10.32$\\
vuds\_cosmos\_510581738 & $4.50$ & $ -1.99$ & $<11.51$ & $10.38$ & $10.69$\\
vuds\_cosmos\_510596653 & $4.57$ & $ -99.00$ & $<11.59$ & -- & --\\
vuds\_cosmos\_510605533 & $4.51$ & $ -2.28$ & $<11.49$ & -- & $10.55$\\
vuds\_cosmos\_510786441 & $4.46$ & $ -2.00$ & $<11.40$ & $10.94$ & $11.26$\\
vuds\_cosmos\_5110377875 & $4.54$ & $ -1.26$ & $<11.62$ & $11.58$ & $11.57$\\
vuds\_cosmos\_5131465996 & $4.46$ & $ -1.79$ & $<11.42$ & $10.35$ & $10.50$\\
vuds\_cosmos\_5180966608 & $4.53$ & $-0.83$ & $11.75$ & $11.58$ & $11.50$\\
vuds\_efdcs\_530029038 & $4.42$ & $-1.49$ & $11.49$ & $11.25$ & $11.30$\\
\end{longtable}


\end{appendix}

\end{document}